\documentstyle[11pt,aasms4,epsf]{article}

\begin{document}

\title{Measuring the galaxy power spectrum with multiresolution
decomposition -- III. velocity field analysis}

\author{XiaoHu Yang\altaffilmark{1,2}\quad Long-Long
Feng\altaffilmark{1,2,3}\quad YaoQuan Chu\altaffilmark{1,2}\quad
Li-Zhi Fang\altaffilmark{4}}

\altaffiltext{1}{Center for Astrophysics, University of Science
and Technology of China, Hefei, Anhui 230026,P.R.China}
\altaffiltext{2}{National Astronomical Observatories, Chinese
Academy of Science, Chao-Yang District, Beijing, 100012, P.R.
China} \altaffiltext{3}{Institute for Theoretical Physics,
Academic of Science, Beijing, 100080, P.R. China
}\altaffiltext{4}{Department of Physics, University of Arizona,
Tucson, AZ 85721}

\begin{abstract}

In this paper, we develop the method of analyzing the velocity
field of cosmic matter with a multiresolution decomposition. This
is necessary in calculating the redshift distortion of power
spectrum in the discrete wavelet transform (DWT) representation.
We show that, in the DWT analysis, the velocity field can be
described by discrete variables, which are given by assignment of
the number density and velocity into the DWT modes. These DWT
variables are complete and not redundant. In this scheme, the
peculiar velocity and pairwise velocity of galaxies or particles
are given by field variables. As a consequence, the velocity
dispersion (VD) and pairwise velocity dispersion (PVD) are no
longer measured by number-counting or pair-counting statistic,
but with the ensemble of the field variables, and therefore, they
are free from the bias due to the number-counting and
pair-counting. We analyzed the VD and PVD of the velocity
fields given by the N-body simulation for models of the SCDM,
$\tau$CDM and $\Lambda$CDM. The spectrum (scale-dependence) of
the VD and PVD show that the length scale of the two-point
correlation of the velocity field is as large as few tens
h$^{-1}$ Mpc. Although the VD and PVD show similar behavior in
some aspects, they are substantially different from each other.
The VD-to-PVD ratio shows the difference between the
scale-dependencies of the VD and PVD. More prominent difference
between the VD and PVD is shown by probability distribution
function. The one-point distribution of peculiar velocity is
approximately exponential, while the pairwise velocity's is
lognormal, i.e. of long tail. This difference indicates that the
cosmic velocity field is typically intermittent.

\end{abstract}

\keywords{cosmology: theory - large-scale structure of universe}

\section{Introduction}

The radial (redshift) distance of galaxies is distorted by their
peculiar velocity. Observation of redshift survey provides only a
distorted picture of the galaxy spatial distribution in the
radial direction. The galaxy power spectrum $P^S(k)$ in the
Fourier representation is measured from the redshift-distorted
galaxy distribution. It is the power spectrum in redshift space. To
recover the power spectrum in real space, $P^R(k)$, one needs to map the
distributions of galaxies from the redshift space to the real space.
Semi-phenomenological theory of redshift distortion show that the
mapping between redshift and real spaces gives
\begin{equation}
P^S(k)=G(k)P^R(k),
\end{equation}
where redshift distortion factor $G(k)$ depends on two parameters:
redshift distortion parameter $\beta$, and 1-D peculiar velocity
dispersion (VD) $\sigma^v$. This leads to two effects:
1. the enhancing of power on large scales due to the linear effect of
redshift distortion; 2. the suppressing of power on small scales due
to the random motions of galaxies inside virialized groups and
clusters of galaxies.

In deriving the relation (1), the following assumptions are employed:
1. the effect of the coupling between the peculiar velocity and the
density perturbations is linear (Kaiser 1987); 2. the effect of the random
motions of galaxies is independent of the density perturbations of
cosmic mass field, 3. the probability distribution function (PDF)
of random peculiar velocity or the pairwise velocity is either
gaussian or exponential. Therefore, besides the linear effect, or
point 1, eq.(1) does not require the information of the spatial
distribution of peculiar velocity of galaxies (or dark matter).
The parameter $\sigma^v$ actually is measured from the statistics of
number-counting of galaxies or pair-counting of galaxies. That is,
the spatial distribution of the peculiar velocity of galaxies or
cosmic matter, ${\bf v}({\bf x})$, is not treated as a random field.

Recently, we developed the method of measuring the galaxy power
spectrum with a space-scale (multiresolution) decomposition, i.e.
measuring power spectrum in the representation of discrete wavelet
transform (DWT) (Fang \& Feng 2000, Yang et al. 2001).
Unlike eq.(1),  the redshift distortion of the
DWT power spectrum must be described by the statistics of the
velocity field ${\bf v}({\bf x})$, as the number-counting or
pair-counting statistics are not enough. For instance, the redshift
distortion of the power of a DWT mode $\psi({\bf x})$ is dependent on
the projection of the velocity field on this mode, i.e.
$\int {\bf v}({\bf x})\psi({\bf x})d{\bf x}$, which cannot be
simply measured by number-counting or pair-counting of galaxies.

The purpose of this paper is to develop the method of analyzing
the velocity field of galaxies of dark matter particles as a
random field. That is, we introduce the variables of the velocity
field by an orthogonal DWT decomposition of the field, and
calculate all statistics of  velocity field by average over the
ensemble of the field variables. With this field description, we
study the problems referring eq.(1): 1.) the scale dependence of
velocity dispersion (VD) and pairwise velocity dispersion (PVD);
2.) the PDF of peculiar velocity and pairwise velocity; 3.) the
local correlation between the velocity field and density field of
cosmic matter.

Other motivation of studying velocity field  is from the
intermittency of cosmic density field. The analysis on high
resolution data of QSO's Ly$\alpha$ forests has revealed that the
cosmic mass field is significantly intermittent (Jamkhedkar, Zhan
\& Fang 2000; Feng, Pando \& Fang 2001; Zhan, Jamkhedkar \&
Fang 2001). That is, the PDF of local density fluctuations on small
scales is neither gaussian nor exponential, but long-tailed.
Taking into account the coupling between the local density fluctuations
and peculiar velocity, we can expect that the PDF of the velocity
field should be neither gaussian nor exponential, but long-tailed
on small scales. This study also requires a decomposition by
orthogonal basis, as a superposition of independent variables
may erase the long-tail feature due to the central limit theorem.

The paper will be organized as follows. \S 2 introduces the multiresolution
analysis of the velocity field. It will focus on the DWT description of the
velocity dispersion (VD) and pairwise velocity dispersion (PVD). \S 3
develops the method of analyzing the scale-dependence of the VD and PVD,
and demonstrate it with an uniformly random sample. \S 4 studies the VD and PVD
of N-body simulated samples, including the scale dependence of the VD and
PVD, the  VD-to-PVD ratio, and the PDFs of the peculiar
velocity and pairwise velocity. In \S 5, we show the correlation between
the VD, PVD and the local density and local density fluctuations. Finally,
the conclusions and discussions will be in \S 6.

\section{Velocity fields in the DWT representation}

\subsection{Peculiar velocity dispersion and relative velocity
   dispersion}

For a given galaxy catalog, one can write down a number density distribution
of galaxies as
\begin{equation}
n({\bf x})=\sum_{n=1}^{N_g}w_n\delta^D({\bf x-x}_n),
\end{equation}
where $N_g$ is the total number of galaxies in the catalog, ${\bf x}_n$ is
the position of the $n$th galaxy, $w_n$ is its weight, and $\delta^D$ the 3-D
Dirac $\delta$ function. $n({\bf x})$ is a sampling of the cosmic mass density
distribution.

The velocity field, ${\bf v}({\bf x})$, describes the deviation of
 particle motion from Hubble flow. The peculiar velocity of
galaxies, ${\bf v}({\bf x}_n)$ and $n=1...N_g$, is a  sampling of
the velocity field. Thus the continuous velocity field ${\bf
v}({\bf x})$ can be constructed by the convolution of the particle
velocity with an assignment function $W(\eta)$,
\begin{equation}
{\bf v}_r({\bf x})=
  \frac{\sum_{n=1}^{N_g} W(|{\bf x_n-x}|/r)w_n{\bf v}({\bf x}_n)}
  {\sum_{n=1}^{N_g} W(|{\bf x_n-x}|/r)w_n}
\end{equation}
where $r$ is a filter scale. One of the simplest example of assignment
functions is the top-hat window function, i.e. $W(\eta)=1$ for
$\eta \leq 1/2$ and 0 for $\eta > 1/2$. Thus, ${\bf v}_r({\bf
x})$ can be considered as a sampling of the cosmic velocity field
${\bf v}({\bf x})$ smoothed on the scale $r$.

Using ${\bf v}_r({\bf x})$, the velocity dispersion (VD) is defined as
the variance of the one-point distribution of the random variable
${\bf v}_r({\bf x})$, i.e.
\begin{equation}
[\sigma^v_r({\bf x})]^2 = \langle {\bf v}^2_r({\bf x}) \rangle.
\end{equation}
where $ \langle ...\rangle$ means the ensemble average.

Using eq.(3), one can define the relative velocity by
\begin{equation}
\Delta {\bf v}_r({\bf x})={\bf v}_{r/2}[{\bf x+(r/2)}]-
  {\bf v}_{r/2}[{\bf x- (r/2)}],
\end{equation}
where $|{\bf r}|=r$. Obviously, $\Delta {\bf v}_r({\bf x})$
measures the difference of the velocity field at ${\bf x+r/2}$
and ${\bf x-r/2}$. If $r$ is small, and there is only one galaxy
${\bf x_n}$ in the sphere $r/2$ at ${\bf x+(r/2)}$, and one galaxy
${\bf x_m}$ in the sphere $r/2$ at ${\bf x-(r/2)}$, we have
$\Delta {\bf v}_r({\bf x}) = {\bf v}({\bf x_n}) - {\bf v}({\bf
x_m})$. Therefore, the measure eq.(5) contains all information of
the relative velocity of individual galaxy pairs.

Yet, the relative velocity defined by eq.(5) is not completely the
same as pairwise velocity of galaxies. Usually, the pairwise velocity
is defined as $v_{12} -\overline{v_{12}}$, where $v_{12}$ is the
relative velocity of galaxy pair 1 and 2 along line-of-sight, and
$\overline{v_{12}}$ is the mean relative velocity, or the in-fall
motion, which generally is modeled by a similarity solution (Davis
\& Peebles 1983). Therefore, the measure of pairwise velocity is
model-dependent.

The reason of using the relative velocity eq.(5) mainly is that the
variables $\Delta {\bf v}_r({\bf x})$ are able to provide an
information-lossless and model-independent decomposition of the velocity
field (\S 2.2). Moreover, the in-fall term of the pairwise velocity
measure is negligible for small scales. Therefore, at least on small
scales, the definition eq.(5) is a reasonable generalization of the
pairwise velocity of galaxies on small scales. We may still call
$\Delta {\bf v}_r({\bf x})$ the variables of pairwise velocity
decomposition.

Thus, the pairwise velocity dispersion (PVD) is defined as the variance
of the one-point distribution of the random variable
$\Delta {\bf v}_r({\bf x})$, i.e.
\begin{equation}
[\sigma^{pv}_r({\bf x})]^2 =
 \langle [\Delta {\bf v}_r({\bf x})]^2 \rangle.
\end{equation}

\subsection{Need for scale-space decomposition}

A common feature of ${\bf v}_r({\bf x})$ [eq.(3)] and $\Delta
{\bf v}_r({\bf x})$ [eq.(5)] is that each quantity depends on two
variables: the position ${\bf x}$ and the scale $r$. That is, the
VD and PVD measure the random velocity field under a space-scale
decomposition. This point can be seen more clearly by rewriting
eqs.(3) and (5) using eq.(2). We have
\begin{equation}
{\bf v}_r({\bf x}) =
  \frac{\int n({\bf x'}){\bf v}({\bf x'})W(|{\bf x'-x}|/r)d{\bf x'}}
  {\int n({\bf x'})W(|{\bf x'-x}|/r)d{\bf x'}}.
\end{equation}
and
\begin{equation}
\Delta {\bf v}_r({\bf x})
   =\frac{\int n({\bf x'}){\bf v}({\bf x'})U_r({\bf x'-x})d{\bf x'}}
    {\int n({\bf x'})W(|{\bf x'-x}|/r)d{\bf x'}},
\end{equation}
where function $U({\bf x'-x})$ is given by the subtractions
of window functions, i.e.
\begin{equation}
U_r({\bf x'-x})=W[|{\bf x'-x+r/2}|/(r/2)]-W[|{\bf x'-x - r/2}|/(r/2)],
\end{equation}
or
\begin{equation}
U_r({\bf x'-x})=\left\{
\begin{array}{ll}
  1 & \mbox{ $|{\bf x' -x -r/2}| <r/2 $} \\
 -1 & \mbox{ $|{\bf x' - x + r/2}| <r/2 $} \\
  0 &
\mbox{otherwise.}
\end{array} \right.
\end{equation}
Function $U_r({\bf x-x'})$ is similar to the Haar wavelet, of which in
1-D is
\begin{equation}
\psi(x)=\left\{
\begin{array}{ll}
  1 & \mbox{ $0 \leq x \leq 1/2$} \\
 -1 & \mbox{ $1/2 < x \leq 1 $} \\
  0 &
\mbox{otherwise.}
\end{array} \right.
\end{equation}
Therefore, the pairwise velocity eq.(8) actually is a decomposition
of the velocity field with a Haar wavelet-like function.

However, the functions $U_r({\bf x'-x})$ with respect to ${\bf x}$
and $r$ don't exhibit  completeness and
orthogonality. Therefore, the decomposition with $U_r({\bf
x'-x})$ may lead to loss of  information if $U$'s are incomplete, or
cause false correlations if they are redundant. To have a proper
measure of the VD and PVD, we call on discrete wavelet transform (DWT),
which provides a complete and unredundant space-scale
decomposition. The first approach of describing cosmic velocity
field by wavelet is given by Rauzy, Lachieze-Rey \& Henriksen
(1993). However, they use continuous wavelets, which give rise to
a redundant decomposition (Farge 1992).

\subsection{The DWT variables of velocity field}

For the details of the mathematical properties of the DWT refers
to Mallat (1989a,b); Meyer (1992); Daubechies, (1992), and for
physical applications, refer to Fang \& Thews (1998). The Haar
wavelet [eq.(12)] provides a clear picture of the DWT
decomposition, and it is also easy for numerical work. However,
the Haar wavelet is discontinues, and therefore, it is not well
localized in scale space. To our work, the most important
properties of the basis for the scale-space decomposition are 1.)
orthogonality, 2.) completeness, and 3.) locality in both scale
and physical spaces. Therefore, wavelets with compactly supported
basis are suitable to the velocity field analysis. Among the
compactly supported orthogonal basis, the Daubechies 4 (D4) is
easy for numerical calculation. We will use wavelet D4. The basic
scaling function $\phi(x)$, basic wavelet $\psi(x)$ and their
Fourier transform of the D4 are shown in Fig. 1.

To simplify the notation, we consider an 1-D density field $n(x)$
and velocity field $v(x)$ on spatial range $L$. The result is
straightforward generalized to 3-D fields in \S 3.3.

For doing the DWT analysis, the space
$L$ is chopped into $2^j$ segments labelled by $l=0,1,...2^j-1$. Each
of the segments has size $L/2^j$. The index $j$ can be a positive integral.
It stands for scale $L/2^j$. The index $l$ is for position, and
corresponds to spatial range $lL/2^j < x < (l+1)L/2^j$.

For a random field $n(x)$, the DWT analysis is performed by the scaling
functions $\phi_{j,l}(x)=(2^j/L)^{1/2}\phi(2^j/L-l)$, and wavelets
$\psi_{j,l}(x)=(2^j/L)^{1/2}\psi(2^j/L-l)$. The scaling
functions play the role of window function. Generally, $\phi_{j,l}(x)$ is
a window in the segment $l$. They are used to
calculate the mean field in the segment $l$. The wavelets $\psi_{j,l}(x)$
essentially is similar to the function $U$ of eq.(10). They are used to
extract the fluctuations of the fields at the segment $l$, i.e. they are
 used to calculate the difference
between the mean fields at space ranges $lL/2^j < x < (l+1/2)L/2^j$ and
$(l+1/2)L/2^j < x < (l+1)L/2^j$.

The scaling functions and wavelets $\psi_{j,l}(x)$ satisfy the
orthogonal relations as
\begin{equation}
\int \phi_{j,l}(x)\phi_{j,l'}(x)dx= \delta_{l,l'},
\end{equation}
\begin{equation}
\int \psi_{j,l}(x)\psi_{j',l'}(x)dx=\delta_{j,j'} \delta_{l,l'},
\end{equation}
\begin{equation}
\int\phi_{j,l}(x)\psi_{j',l'}(x)dx =0, \ \ \ \mbox{if $j'\geq j$}.
\end{equation}

With these properties, a 1-D random field $n(x)$ can be decomposed into
\begin{equation}
n(x) = n^{j}(x) + \sum_{j'=j}^{\infty} \sum_{l=0}^{2^{j'}-1}
  \tilde{\epsilon}^n_{j',l} \psi_{j',l}(x),
\end{equation}
where
\begin{equation}
n^j(x)=\sum_{l=0}^{2^j-1}\epsilon^n_{j,l}\phi_{j,l}(x).
\end{equation}
The scaling function coefficient (SFC) $\epsilon_{j,l}^n$ and the
wavelet function coefficient (WFC), $\tilde{\epsilon}_{j,l}^n$
are given by
\begin{equation}
\epsilon^n_{j,l} =\int n(x)\phi_{j,l}(x)dx,
\end{equation}
and
\begin{equation}
\tilde{\epsilon}^n_{j,l}=\int n(x)\psi_{j,l}(x)dx,
\end{equation}
respectively. The SFC $\epsilon^n_{j,l}$ measures the mean
$n(x)$ in the segment $l$, while the WFC
$\tilde{\epsilon}^n_{j,l}$ measures the fluctuations of field
$n(x)$ at $l$ on scale $j$.

The first term on the r.h.s. eq.(15), $n^{j}(x)$, is the field
$n(x)$ smoothed on the scale $j$, while the second term contains
all information on scales $\geq j$. Because of the orthogonal
relation eq.(14), the decomposition between the scales of $ <j$ (first
term) and $\geq j$ (second term) in eq.(15) is unambiguous. The
value of $j$
in eq.(15) can be any integer, and therefore, a scale-by-scale
decomposition becomes possible.

Since scale $r$, position $x$, and window function $W(|{x'-x}|/r)$
of eq.(3) correspond to, respectively, $j$, $l$, and $\phi_{j,l}(x')$
of the DWT analysis, the DWT counterpart of eq.(3) (for 1-D) is
\begin{equation}
  v_{j,l} = \frac{\int v(x)n(x)\phi_{j,l}(x) dx}
                 {\int n(x)\phi_{j,l}(x) dx}=
          \frac{\epsilon^v_{j,l}}{\epsilon^n_{j,l}}
\end{equation}
where $\epsilon^v_{j,l}$ and $\epsilon^n_{j,l}$ are respectively,
the SFC of field $v(x)n(x)$ and $n(x)$, i.e.
\begin{equation}
\epsilon^v_{j,l}=\int v(x)n(x)\phi_{j,l}(x) dx =
  \sum_{n=1}^{N_g}w_nv(x_n)\phi_{j,l}(x_n)
\end{equation}
and
\begin{equation}
\epsilon^n_{j,l}=\int n(x)\phi_{j,l}(x) dx =
  \sum_{n=1}^{N_g}w_n\phi_{j,l}(x_n).
\end{equation}
$v_{j,l}$ is the mean
velocity in the spatial range $lL/2^j < x < (l+1)L/2^j$.

Similarly, $U_r$ corresponds to $\psi_{j,l}$, and therefore, the
DWT counterpart of eq.(5) is
\begin{equation}
\Delta v_{j,l} = \frac{\int v(x)n(x)\psi_{j,l}(x) dx}
                 {\int n(x)\phi_{j,l}(x) dx}=
       \frac{\tilde{\epsilon}^v_{j,l}}{\epsilon^n_{j,l}},
\end{equation}
where $\tilde{\epsilon}^v_{j,l}$ is the WFC of field $v(x)n(x)$,
i.e.
\begin{equation}
\tilde{\epsilon}^v_{j,l}=\int v(x)n(x)\psi_{j,l}(x) dx =
  \sum_{n=1}^{N_g}w_nv(x_n)\psi_{j,l}(x_n).
\end{equation}
$\Delta v_{j,l}$ is the difference between the mean
velocities of spatial ranges $lL/2^j < x < (l+1/2)L/2^j$ and
$(l+1/2)L/2^j < x < (l+1)L/2^j$.

$v_{j,l}$ and $\Delta v_{j,l}$ are the variables of the velocity
field $v(x)$ in the DWT representation. These variables give a
complete description of the field $v(x)$ without loss of
information. The orthogonality of scaling functions and wavelets
insure that the decomposition does not cause false correlation
among these variables.

It has been pointed out that the galaxy pairwise velocity
dispersion measured by conventional techniques, i.e.
pair-counting statistics, is biased by densest regions,
as the statistic actually is pair-weighted (Strauss, Ostriker \&
Cen 1998.) In the language of the velocity field decomposition,
the conventional method is redundant for modes at the dense
regions. In the DWT decomposition, each mode $(j,l)$ corresponds
to a cell in phase space (scale $L/2^j$ and position $l$), and
is described by one variable $v_{j,l}$ or $\Delta v_{j,l}$. That
is, all modes are treated equal. Each degree of freedom of the
velocity field is represented by one variable, regardless the
number of galaxies in the cell. Therefore, the description of
$v_{j,l}$ and $\Delta v_{j,l}$ is free from the bias of
pair-weight.

\section{The VD and PVD spectrum of a velocity field}

\subsection{The VD and PVD spectrum}

With eqs.(19) and (22), the VD and PVD are given by ensemble
averages of $v_{j,l}^2$ and $\Delta v_{j,l}^2$, i.e.
\begin{equation}
\sigma^v_{j,l}=\langle v_{j,l}^2 \rangle ^{1/2}=
\left \langle \left [\frac{\int v(x)n(x)\phi_{j,l}(x) dx}
                 {\int n(x)\phi_{j,l}(x) dx}  \right ]^2
      \right \rangle ^{1/2}
\end{equation}
and
\begin{equation}
\sigma^{pv}_{j,l}=\langle \Delta v_{j,l}^2 \rangle ^{1/2}=
\left \langle \left [\frac{\int v(x)n(x)\psi_{j,l}(x) dx}
                 {\int n(x)\phi_{j,l}(x) dx}  \right ]^2
      \right \rangle ^{1/2}.
\end{equation}
If the field is statistically uniform, $\sigma^v_{j,l}$ and
$\sigma^{pv}_{j,l}$ are independent of $l$. One can define the VD
and PVD on scale $j$ by $\sigma^v_j \equiv \sigma^v_{j,l}$, and
$\sigma^{pv}_j \equiv \sigma^{pv}_{j,l}$, which are referred to as
the VD and PVD spectrum of the velocity field, respectively.

If the ``fair sample hypothesis" (Peebles 1980) holds, the ensemble
average $\langle ...\rangle$ can be replaced by a spatial average
over $l$. Eqs.(24) and (25) become
\begin{equation}
\sigma^v_j =
  \left [  \frac{1}{2^j} \sum_{l=0}^{2^j-1} v_{j,l}^2 \right ]^{1/2}
 = \left [ \frac{1}{2^j} \sum_{l=0}^{2^j-1}
   \left ( \frac{\epsilon^v_{j,l}}{\epsilon^n_{j,l}} \right )^2
   \right ]^{1/2}.
\end{equation}
and
\begin{equation}
\sigma^{pv}_j =\left [\frac{1}{2^j} \sum_{l=0}^{2^j-1}\Delta v_{j,l}^2
   \right ]^{1/2}= \left [\frac{1}{2^j} \sum_{l=0}^{2^j-1}
   \left( \frac{\tilde{\epsilon}^v_{j,l}}{\epsilon^n_{j,l}}\right )^2
   \right ]^{1/2}.
\end{equation}
Eqs.(26) and (27) show that the VD and PVD of a velocity field are
given by the SFCs and WFCs of the DWT variables, respectively.
For a non-gaussian random field $F(x)$, the statistical behaviors of
the field quantity $F(x)$ and its spatial difference $F(x)-F(x+r)$
generally are different.  The statistical behaviors of the VD and PVD
are different from each other.

As a useful variant of eq.(24), we define a modified VD by
\begin{equation}
\bar{\sigma}_j^v =
   \left [ \frac{\langle [\int v(x)n(x)\phi_{j,l}(x) dx]^2 \rangle }
        {\langle [ \int n(x)\phi_{j,l}(x) dx  ]^2 \rangle } \right ]^{1/2}.
\end{equation}
Obviously, if the distribution $n(x)$ is randomly uniform,
$\bar{\sigma}_j^v$ will be the same as $\sigma_j^v$. If velocity
field is long-range correlated, i.e.
$\langle v(x)v(x') \rangle \simeq$ const, and no correlation with
$n(x)$, we have also $\bar{\sigma}_j^v \simeq \sigma_j^v$, even
$n(x)$ is not uniform.

Because $\langle [\int v(x)n(x)\phi_{j,l}(x) dx ]^2 \rangle=
 (1/2^j)\sum (\epsilon^{v}_{\bf j,l})^2$ and
$\langle [ \int n(x)\phi_{j,l}(x) dx]^2 \rangle=(1/2^j)\sum
   (\epsilon^n_{\bf j,l})^2$,
eq.(28) yields
\begin{equation}
\bar{\sigma}_j^v =
  \left [ \frac{\sum_{l=0}^{2^j-1} (\epsilon^{v}_{\bf j,l})^2}
             {\sum_{l=0}^{2^j-1}(\epsilon^n_{\bf j,l})^2} \right  ]^{1/2}.
\end{equation}

Similarly, a modified PVD can be defined by
\begin{equation}
\bar{\sigma}_j^{pv} = \left [ \frac{\sum_{l=0}^{2^j-1}
                  (\tilde{\epsilon}^{v}_{\bf j,l})^2}
             {\sum_{l=0}^{2^j-1}
                  (\epsilon^n_{\bf j,l})^2} \right ]^{1/2}.
\end{equation}
The difference between $\sigma^v_j$ and $\bar{\sigma}^v_j$, or
$\sigma^{pv}_j$ and $\bar{\sigma}^{pv}_j$ can be used as an
indicator of the correlations between velocity and density fields
(\S 5).

\subsection{An example: random fields}

To demonstrate the VD and PVD spectra, let's consider 1-D random
fields, which are produced by uniformly random distribution of
$N$ particles in range $L$. Each particle $n$ is assigned a
velocity $v_n$ drawn from a Gaussian distribution of the
velocity with zero mean and variance 300 km s$^{-1}$. The two-point
correlation
function of number density $\langle n(x)n(x') \rangle =
\bar{n}^2=(N/L)^2$. The two-point velocity correlation function of
velocity field $\xi^v(x-x')\equiv \langle v(x)v(x')\rangle$ would be
\begin{equation}
\xi^v(x-x') = \left\{
\begin{array}{ll}
  (300)^2 \ \ \mbox{\ \ km$^2$ s$^{-2}$} & \mbox{ $|x-x'| \simeq L/N$ } \\
  0 &  \mbox{ $|x-x'| > L/N$ .}
\end{array} \right.
\end{equation}

In this case, $\sigma_j^v \simeq \bar{\sigma}_j^v$. Figure 1(a) displays
the modified VD spectrum $\bar{\sigma}^v_j$ of
the random samples with $L=1$ and $N=$ 10$^3$, $10^4$ and 10$^5$.
Fig. 1(a) shows that $\bar{\sigma}^v_j$ increases with $j$, and
saturates at 300 km/s. This behavior can be deduced analytically.
In fact, following eq.(28) and $\langle n(x)n(x') \rangle=$
const, we have
\begin{eqnarray}
 (\bar{\sigma}^v_{j})^2 &  = &
 \frac{\int \int \langle n(x)n(x')\rangle \langle v(x)v(x')\rangle
   \phi_{j,l}(x)\phi_{j,l}(x') dx dx'}
 {\int \int \langle n(x)n(x')\rangle \phi_{j,l}(x)\phi_{j,l}(x') dx dx'}
 \\ \nonumber
 & = &  \frac{\int \int \xi^v(x-x') \phi_{j,l}(x)\phi_{j,l}(x')dx dx'}
    {\int \int \phi_{j,l}(x)\phi_{j,l}(x') dx dx'}
\end{eqnarray}
Considering that both $\phi_{j,l}(x)$ and $\phi_{j,l}(x')$ are
window-like functions in the spatial range $lL/2^j < x,x' <
(l+1)L/2^j$, the integral
 of the denominator of eq.(31) is proportional to area
$L/2^j \times L/2^j$. On the other hand, the numerator is proportional to
area $L/2^j \times L/N $, the factor $L/N$ is  due to the condition
$|x-x'| \simeq L/N$ in eq.(31). Thus, the VD of eq.(32) is
\begin{equation}
\bar{\sigma}^v_j \simeq \left\{
\begin{array}{ll}
 2^{j/2} 300/\sqrt{N} \mbox{\ \ km$^2$ s$^{-2}$}  &
   \mbox{if $j \ll \ln_2 N$ }\\
 300 \mbox{\ \ km$^2$ s$^{-2}$} & \mbox{if $j\simeq \ln_2 N$ }
\end{array} \right.
\end{equation}
This is just what as indicated in Fig. 1(a).

Figure 1(b) gives the modified PVD spectrum calculated by eq.(30). It
shows the same as the VD spectrum, i.e. $\bar{\sigma}^{pv}_j =
\bar{\sigma}^v_j$. This is expected. For the VD eq.(29), the SFC
$\epsilon^v_{j,l}$ is given by the addition of the velocities of
galaxies within the cell $(j,l)$, while for the PVD eq.(30), the
WFC $\tilde{\epsilon}^v_{j,l}$ is given by the difference between
the velocities of galaxies in $Ll/2^{j+1}$ and
$L(l+1/2)/2^{j+1}$. However, the one point function of the
velocity is Gaussian, the probability of positive and negative
velocity is the same, and therefore, the statistical properties
of the SFC and WFC are the same. Consequently, the VD and PVD
spectra coincide for this random field.

 From Figs. 1(a) and 1(b), we can conclude that if the velocities of
galaxies are
uncorrelated on scales larger than $j$, the VD and PVD spectra will
decrease systematically with the $j$ decreasing by a law $2^{j/2}$ (1-D).
This result is useful to estimate the scale of the correlation length of
velocity field of galaxies.

\subsection{3-D velocity field}

For 3-D velocity field ${\bf v}({\bf x})$, ${\bf x}=(x^1,x^2,x^3)$,
the DWT decomposition is based on the 3-D scaling functions and wavelets,
which can be constructed by a direct product of 1-D wavelet basis
(Fang \& Thews 1998). For instance, in a 3-D volume
$L_1\times L_2 \times L_3$, we have scaling function as
\begin{equation}
\phi_{\bf j,l}({\bf x})=\phi_{j_1,l_1}(x_1)\phi_{j_2,l_2}(x_2)
 \phi_{j_3,l_3}(x_3),
\end{equation}
and wavelets as
\begin{equation}
\psi_{\bf j,l}({\bf x})=\psi_{j_1,l_1}(x_1)\psi_{j_2,l_2}(x_2)
 \psi_{j_3,l_3}(x_3),
\end{equation}
where ${\bf j}=(j_1,j_2,j_3)$, ${\bf l}=(l_1,l_2,l_3)$, and
$l_i=0...2^{j_i-1}$, $i=1,2,3$. The wavelet $\psi_{\bf j,l}({\bf x})$
is non-zero mainly in a volume
$L_1/2^{j_1} \times L_2/2^{j_2}\times L_3/2^{j_3}$, and around the
position
$(x_1=l_1L_1/2^{j_1}, x_2=l_2L_2/2^{j_2},x_3=l_3L_3/2^{j_3})$, i.e.
they localized in both scale and physical space.

The 3-D generalization of the DWT variables $v_{j,l}$ [eq.(19)] and
$\Delta v_{j,l}$ [eq.(22)] are given by
\begin{equation}
v_{i \ {\bf j,l}} =
   \frac{\epsilon^{v_i}_{\bf j,l}}{\epsilon^n_{\bf j,l}}
\end{equation}
and
\begin{equation}
\Delta v_{i \ {\bf j,l}} =
       \frac{\tilde\epsilon^{v_i}_{\bf j,l}}{\epsilon^n_{\bf j,l}},
\end{equation}
where $\epsilon^{v_i}_{\bf j,l}$ and $\epsilon^{n}_{\bf j,l}$ are
the SFCs as
\begin{equation}
\epsilon^{v_i}_{\bf j,l}=\int n({\bf x})v_i({\bf x})
  \phi_{\bf j,l}({\bf x})d{\bf x}
  =\sum_{n=1}^{N_g}w_n v_i({\bf x}_n) \phi_{\bf j,l}({\bf x}_n),
\end{equation}
\begin{equation}
\epsilon^{n}_{\bf j,l}=\int n({\bf x})\phi_{\bf j,l}({\bf x})d{\bf x}
    =\sum_{n=1}^{N_g}w_n \phi_{\bf j,l}({\bf x}_n),
\end{equation}
and $\tilde{\epsilon}^{v_i}_{\bf j,l}$ is the WFC as
\begin{equation}
\tilde{\epsilon}^{v_i}_{\bf j,l}=
\int v_i({\bf x})\psi_{\bf j,l}({\bf x}) d{\bf x}=
  \sum_{n=1}^{N_g}w_n v_i({\bf x}_n)\psi_{\bf j,l}({\bf x}_n).
\end{equation}
The SFCs $\epsilon^{n}_{\bf j,l}$ [eq.(39)] are assigned at regular
grids $l_i=0...2^{j_i}$ and $i=1,\ 2,\ 3$. That is, $\epsilon^{n}_{\bf j,l}$ is
an assignment of the number density distribution $n({\bf x})$ onto
grids $({\bf j,l})$. Eq.(39) shows that the $n$th galaxy (or particle)
is assigned onto grid ${\bf j,l}$ by weight $\phi_{\bf j,l}({\bf x}_n)$
(Fang \& Feng 2000). Similarly, SFCs $\epsilon^{v_i}_{\bf j,l}$
is an assignment of the distribution $v_i({\bf v})n({\bf x})$ onto
grid ${\bf j,l}$. Therefore, eqs.(36) and (37) essentially are the
same as the DWT mass assignment (Fang \& Feng, 2000). We may call
it the DWT velocity assignment.

 From eqs.(36) and (37), we have
\begin{equation}
\sigma^{v_i}_{\bf j}=[\langle (v_{i\ {\bf j,l}})^2\rangle]^{1/2}
  =\left [\frac{1}{2^{(j_1+j_2+j_3)}}
\sum_{{\bf l}=0}^{2^{\bf j}-1}
 \left (\frac{\epsilon^{v_i}_{\bf j,l}}{\epsilon^{n}_{\bf j,l}} \right )^2
   \right ]^{1/2},
\end{equation}
\begin{equation}
\bar{\sigma}^{v_i}_{\bf j}=\left [\frac{\sum_{{\bf l}=0}^{2^{\bf j}-1}
  (\epsilon^{v_i}_{\bf j,l})^2}
{\sum_{{\bf l}=0}^{2^{\bf j}-1}(\epsilon^{n}_{\bf j,l} )^2}
   \right ]^{1/2}
\end{equation}
for 3-D VD and modified VD, and
\begin{equation}
\sigma^{pv_i}_{\bf j}=[\langle (\Delta v_{i\ {\bf j,l}})^2\rangle]^{1/2}
 =\left [\frac{1}{2^{(j_1+j_2+j_3)}}
\sum_{\bf l=0}^{2^{\bf j}-1}
 \left (\frac{\tilde{\epsilon}^{v_i}_{\bf j,l}}
   {\epsilon^{n}_{\bf j,l}} \right )^2
   \right ]^{1/2},
\end{equation}
\begin{equation}
\bar{\sigma}^{pv_i}_{\bf j}=\left [\frac{
\sum_{\bf l=0}^{2^{\bf j}-1}
  (\tilde{\epsilon}^{v_i}_{\bf j,l})^2}
{\sum_{\bf l=0}^{2^{\bf j}-1}(\epsilon^{n}_{\bf j,l} )^2}
   \right ]^{1/2}
\end{equation}
for 3-D PVD and modified PVD.

\section{Velocity field of simulation samples}

\subsection{Simulation samples}

To demonstrate the velocity field of cosmic matter, we produce ten
realizations of the N-body simulation for each model of the SCDM,
$\tau$CDM and $\Lambda$CDM. The parameters
$(\Omega_0,\Lambda,\Gamma,\sigma_8)$ are taken to be
(1.0,0.0,0.5,0.55) for standard CDM model (SCDM),
(1.0,0.0,0.25,0.55) for a variant of SCDM model ($\tau$CDM) and
(0.3,0.7,0.21,0.85) for a low density flat model ($\Lambda$CDM). We use
modified AP$^3$M code (Couchman, 1991) to evolve $128^3$ cold
dark matter particles in a periodic cube of side length L=
256 h$^{-1}$Mpc. The linear power spectrum is taken from the
fitting formula given in Bardeen et.al. (1986).

In our simulation, the so-called ``glass" configuration is used
to generate the unperturbed uniform distribution of particles,
and the Zel'dovich approximation is then applied to set up the
initial perturbation. For the mass assignment on the grid and the
calculation of the force on a given particle from interpolation
of the grid values, we use the triangular-shaped cloud (TSC)
method. The starting redshift of the simulation is taken to be
$z_i=15$ for the SCDM model and $z_i=25$ for $\Lambda$CDM and $\tau$CDM.
The force softening parameter $\eta $ in the comoving system
decreases with time as $\eta \propto 1/a(t)$. Its initial value
is taken to be $\eta=384$ h$^{-1}$\, kpc and the minimum value
to be $\eta_{\min}=128$ h$^{-1}$\, kpc, corresponding to 15\% and
5\% of the grid size respectively. We use the ``leap-frog''
scheme in the single-step integration of time evolution, and take
600 total integration steps down to $z=0$.

\subsection{The VD and PVD spectra of the simulation samples}

Now we calculate the VD spectrum of the simulation samples by
eqs.(41) and (42), and the PVD spectrum by eqs.(43) and (44). A
problem in using eqs.(41) and (43) is the ambiguity caused by
extremely low values of galaxy number density in voids or
underdense regions in the galaxy distribution, i.e. $n({\bf
x})\simeq 0$. Within these areas, the corresponding SFCs
$\epsilon^{n}_{\bf j,l}$ would be very small, even zero, and
hence, lead to very large terms of $\epsilon^{v_i}_{\bf
j,l}/\epsilon^{n}_{\bf j,l}$. These terms are unacceptable,
because in these areas the number density $n({\bf x})$ actually
are dominated by Poisson fluctuations. To control these terms, we
use the so-called ``off-counting'' algorithm, of which the
details are given in Appendix.

Since the cosmic velocity field is isotropic, the VD or PVD for
the three components of $v^i$ and $pv^i$ $i=1,2,3$ should be the
same, i.e.
\begin{equation}
\sigma^{v_1}_{\bf j} = \sigma^{v_2}_{\bf j} = \sigma^{v_3}_{\bf j},
 \hspace{1cm}
\sigma_{\bf j}^{pv_1} =\sigma_{\bf j}^{pv_2} =\sigma_{\bf j}^{pv_3}.
\end{equation}
Therefore, it is enough to calculate one component among $v_i$ and
$pv_i$.

We consider diagonal mode, i.e. $j_1=j_2=j_3=j$. In this case, the
VD, $\sigma^{v_i}_j$, and PVD, $\sigma^{pv_i}_j$, depend only on
one scale parameter $j$. The physical scale  of mode $j$ is
$256/2^j$ h$^{-1}$ Mpc, or wavenumber $k= 2^{j+1}\sqrt{3}\pi/256$.
Figures 3 and 4 display, respectively, the $j$- or $k$-dependence
of $\sigma^{v_i}_j$ and $\bar{\sigma}^{v_i}_j$ for models of SCDM,
$\tau$CDM and $\Lambda$CDM. In comparison, the figures also plot the VD
spectra for two types of random samples, which are given by
assigning $N_g$ velocities to randomly distributed particles
(hereafter random1), and randomizing the assignment of the $N_g$
velocities with particles ${\bf x}_n$ ($n=1...N_g$) (hereafter
random2).

Figures 5 and 6 give, respectively, the $j$- or $k$-dependence of
$\sigma^{pv_i}_j$ and $\bar{\sigma}^{pv_i}_j$ of the simulation
and random samples for models of SCDM, $\tau$CDM and $\Lambda$CDM.

As mentioned in \S 3.2, if the peculiar velocities are not
correlated on scales larger than $j$, or $\xi^v({\bf x-x'})
\simeq 0$ for $|{\bf x-x'}| > L/2^j$, the VD and PVD spectra will
decrease with the decreasing $j$ according to  the law of
$2^{j/2}$ (1-D) [eq.(33)],
or $2^{3j/2}$ (3-D). The spectra of random samples of Figs. 3
and 5 are decreasing with $j$ as the $2^{3j/2}$ law. However, the
VD spectra of the simulation samples are almost flat, i.e.
scale-independent, in the range of $k > 0.2$ h Mpc$^{-1}$, and
slightly decreasing for $k < 0.2$ h Mpc$^{-1}$. This result
shows that the peculiar velocities are spatially correlated
on scales at least few 10 h$^{-1}$ Mpc. This is probably caused by
the in-fall process of forming halos on scales 10 and 20 h$^{-1}$
Mpc (Xu, Fang \& Wu 2000.)

The PVD spectrum decreases with $j$ on scale $k \simeq 1$
h Mpc$^{-1}$, but it is slower than the $2^{3j/2}$ law.
Accordingly, the correlation of the PVD on large scales is also
significant.

Recall that in the recovery of the real space Fourier power
spectrum from galaxy redshift surveys, to account for the
redshift distortion effect, it is usually assumed that the
peculiar velocity dispersion $\sigma^v$ is scale-independent. The
DWT analysis show that the assumption of a constant $\sigma^v$ is
reasonable on scales $k > 0.2$ h Mpc$^{-1}$, but may have
uncertainty of factor 2 or 3 on large scales.

Although the VD and PVD of CDM models have similar scale-dependence,
a difference between the VD and PVD can already be seen from Figs.
3 - 6. Figs. 3 and 4 show that for the CDM models, the modified
VD $\bar{\sigma}^{v_i}_j$ is almost the same as $\sigma^{v_i}_j$ on
all scales, while their $\sigma^{pv_i}_j$ are generally
larger than $\bar{\sigma}^{pv_i}_j$, especially on small scales
(Figs. 5 and 6). This result indicates that the peculiar velocity
possesses long self-correlation length, but is less correlated with
local number density.

This conclusion can also be seen from the behavior of random1 and
random2 samples. The only difference between random1 and random2
is that the former has a uniform distribution $n({\bf x})$ while
the latter is inhomogeneous. Figs. 3 - 4 shows that for random1
the modified VD is the same as the VD, but for random2, the
modified VD is substantially lower than the VD on small scales.
The latter is due to the clustering of density field $n({\bf
x})$. The summation ${\frac {1}{2^{\bf j}}}\sum_{{\bf
l}=0}^{2^{\bf j}-1}[1/\epsilon^{n}_{\bf j,l}]^2$ general is
larger than $1/({\frac {1}{2^{\bf j}}}\sum_{{\bf l}=0}^{2^{\bf
j}-1}[\epsilon^{n}_{\bf j,l}]^2)$ for inhomogeneous distribution
$n({\bf x})$. Thus, the result of $\bar{\sigma}^{v_i}_j \simeq
\sigma^{v_i}_j$ for CDM models means that the correlation of
peculiar velocity field is not affected by the inhomogeneity
density field.

\subsection{The VD-to-PVD ratio (Mach number)}

The difference between the VD and PVD can be measured by the
VD-to-PVD ratio as
\begin{equation}
M_j = \frac{\sigma^{v_i}_j}{\sigma^{pv_i}_j}.
\end{equation}
This ratio is essentially to measure the relative strength of the
mean velocity and the fluctuations of velocity field on scale $j$,
i.e. the ratio between the bulk velocity and velocity
fluctuations. Therefore, it is similar to the so-called Mach number
(e.g. Suto \& Fujita 1990).

For the randomized samples, the VD and PVD spectra are identical
(\S 3.2), as both spectra follow the $2^{3j/2}$ (3-D) law. Upon
the definition of eq.(46), the Mach numbers of the random samples
of \S 3.2 are equal to 1 on all scales. For simulation samples,
the VD spectrum is no longer the same as PVD spectrum on all
scales. The deviation of $M_j$ from one is a measure of evolved
velocity field. Generally, $\sigma_j^{v_i}$ is larger than
$\sigma_j^{pv_i}$ on large scales, i.e. the bulk flow is
significantly larger than the PVD. This is due to the in-fall
evolution. But $\sigma_j^{v_i}$ might be is less than
$\sigma_j^{pv_i}$ on small scales.

Figure 7 plots the Mach numbers on scales $j=1 - 7$ for the three
dark matter models. The figure shows that $M_j$ has a peak at $k
\simeq 0.2$ h Mpc$^{-1}$ for SCDM and $\tau$CDM, and $k\simeq
0.1$ for the $\Lambda$CDM. The peak characterizes a typical scale on which
the matter undergoes the in-fall evolution. The clustering in the
$\Lambda$CDM model is earlier than the other two models, and therefore,
the $\Lambda$CDM's peak is on larger scale than the two others. On scales
much larger than the peak, the mass and velocity fields are only slightly
different from a Gaussian field, the VD and PVD are of the same order.
On small scales, $M_j$ becomes small, and finally $\leq 1$, i.e.
the PVD is larger than the VD. This scale-dependence can clearly
be seen with the $\Lambda$CDM model.

It has to be pointed out that the Mach number $M_j$ is a
statistical measure of the entire velocity  field, rather for
individual object. Since the VD and PVD have different
distribution functions (\S 4.5), the Mach number measured from
individual objects might be significantly different from the Mach
number of the entire velocity field.

\subsection{Comparing DWT method with conventional techniques}

It would be interesting to compare the DWT measurement of VD and
PVD with conventional techniques. Usually, the bulk velocity is
given by the amplitude of the $R_s$-smoothed velocity  field
$v_i({\bf x})$ over a volume defined by a normalized
window function $W_R({\bf x})$ of a characteristic scale $R$
\begin{equation}
V_i(R) =\int W_R({\bf x}) v_i({\bf x}) d {\bf x}.
\end{equation}
We calculate the variance $\langle V_i^2(R)\rangle^{1/2}$  with
gaussian window function $W_R \propto  \exp(-r^2/2R^2_s)$ for
SCDM simulation sample. The result is shown in Fig. 8. We also
plot $\sigma^{v_i}_j$ (SCDM) in Fig. 8. The data $\sigma^{v_i}_j$
(SCDM) is the same as Fig. 3. Since the effective radius of the
gaussian window is $R=(3\sqrt{2\pi})^{1/3} R_s$, it corresponds
to $j$ given by $R= (3/4\pi)^{1/3}256/2^j$ h$^{-1}$ Mpc.

Figure 8 shows that the DWT gives exactly the same variance of
bulk velocity as that measured by gaussian smooth.

To compare with the PDV of Fig. 5, we calculate the pairwise
velocity of particles,
$v_{12}(r=|{\bf r}|) =
 |\hat{r}\cdot[{\bf v}({\bf x+r})-{\bf v}({\bf x})]|$,
by pair-counting method. The variance of pairwise velocity,
$\langle v^2_{12}(r) \rangle^{1/2}$, for the SCDM sample is shown in
Fig. 9. The data of $\sigma^{pv_i}_j$ of Fig. 5 is also shown in
Fig. 9. It is interested to see that scales less than 5 h$^{-1}$ Mpc,
we have
$\langle v^2_{12}(r) \rangle^{1/2} \simeq \sqrt{3} \sigma^{pv_i}_j$
and $\langle V^2_{12}(R) \rangle^{1/2} \simeq \sqrt{2}\sqrt{3}
\langle V_i^2(R=r)\rangle^{1/2}$. This result implies that velocity
distribution on these scales is virialized, or quasi-virialized
(Xu, Fang \& Wu 2000).

On scales larger than 5 h$^{-1}$ Mpc,
$\langle v^2_{12}(r) \rangle^{1/2}$ generally is larger than
$\sqrt{3} \sigma^{pv_i}_j$, but less than
$\sqrt{2}\sqrt{3} \langle V_i^2(R)\rangle^{1/2}$. The large value
of $\langle v^2_{12}(r) \rangle^{1/2}$ probably is because
pair-counting measurement is not orthogonal with the
$V_i(R)$ measurement, i.e. the pair-counting techniques may be
biased by $\langle v^2_{12}(r) \rangle^{1/2}$.

The difference between field variable and conventional description
of velocity field can easily be seen in linear regime of the cosmic
clustering. It is well known that the linear relation between the
density fluctuations and peculiar velocity field
$\delta({\bf x})= -(1/H_0\beta)\nabla\cdot {\bf v}({\bf x})$
yields
\begin{equation}
\langle V^2(R)\rangle =\frac {H_0^2 \beta^2}{2\pi^2}
  \int_{0}^{\infty}dk P(k) \hat{W}_R^2(k),
\end{equation}
where $\langle V^2(R)\rangle =\sum_{i=1}^{3}\langle V_i^2(R)\rangle$,
$\beta$ is the redshift distortion parameters, $P(k)$ the
Fourier power spectrum of $\delta({\bf x})$, and $\hat{W}_R(k)$ the
Fourier transform of $W_R({\bf x})$. In the DWT representation, the
$\delta({\bf x})-{\bf v}({\bf x})$ linear relation gives
\begin{equation}
 [\sigma^{v}_j]^2  = \frac {H_0^2\beta^2}{(2\pi)^2L}
 \sum_{n_1,n_2,n_3 = -\infty}^{\infty} \frac{1}{n_1^2 + n_2^2+ n_3^2}
  P(n)|\hat{\phi}(n_1/2^{j})\hat{\phi}(n_2/2^{j})
  \hat{\phi}(n_3/2^{j})|^2,
\end{equation}
where $[\sigma^{v}_j]^2=\sum_{i=1}^{3}[\sigma^{v_i}_j]^2$, the
Fourier power spectrum $P(n)$ is a function of $n$ related to the
wavenumber by $k= 2\pi n/L$, $n^2=n_1^2+n_2^2+n_3^2$, and
$\hat{\phi}$ is the Fourier transform of the basic scaling
function $\phi(n)$ (Fig. 1).

Therefore, the only difference of the DWT measurement of VD
from conventional method is to replace the ordinary window function
$\hat{W}_R^2(k)$ by the DWT window
$|\hat{\phi}(n_1/2^{j})\hat{\phi}(n_2/2^{j})\hat{\phi}(n_3/2^{j})|^2$.
Fig.1 shows that $|\hat{\phi}(n)|^2$ is similar to  ordinary window.
This is why $\langle V^2(R)\rangle \simeq [\sigma^{v}_j]^2$ if
the windows have the same size.

With the linear relation between $\delta({\bf x})$ and ${\bf v}({\bf x})$,
the velocity dispersion within a patch of the velocity field sometimes
is described by
\begin{equation}
\langle \Delta V^2(R)\rangle =\frac {H_0^2 \beta^2}{2\pi^2}
  \int_{0}^{\infty}dk P(k) [1-\hat{W}_R^2(k)].
\end{equation}
The DWT counterpart of eq.(50) is given by the pairwise velocity. It is
\begin{equation}
 [\sigma^{pv}_j]^2 =  \frac {H_0^2\beta^2}{(2\pi)^2L}
  \sum_{n_1,n_2,n_3 = -\infty}^{\infty} \frac{1}{n_1^2 + n_2^2+ n_3^2}
P(n)|\hat{\psi}(n_1/2^{j})\hat{\psi}(n_2/2^{j})\hat{\psi}(n_3/2^{j})|^2,
\end{equation}
where $[\sigma^{pv}_j]^2=\sum_{i=1}^{3}[\sigma^{pv_i}_j]^2$, and
$\hat{\psi}$ is the Fourier transform of basic wavelet $\psi$.

It is clearly to see the difference between eqs.(50) and (51).
The r.h.s. of eq.(50) contains all contribution of $P(k)$ on scale
$kR \geq 1$. On the other hand, Fig. 1 shows that $|\hat{\psi}(n)|^2$
is localized at $n \simeq \pm 1$, and therefore, the r.h.s. of eq.(51)
contains only few terms around $n \simeq \sqrt{3}2^j$. This shows that
eq.(51) is based on unambiguous scale-decomposition.

As for the conventional pair-counting measurement of PVD, one cannot
write down a linear relation between $\langle v^2_{12}(r) \rangle^{1/2}$
and $P(k)$ as eqs.(50) or (51). Although $r$ is used as a scale-indicator,
the pair-counting method rely on a scale-decomposition of the distance $r$
of pairs, but not a scale decomposition of the field. Therefore, the
statistics of $\langle V^2_{12}(R) \rangle^{1/2}$ are different from
$[\sigma^{pv}_j]^2$ based on ensemble of the field variables.

Eqs.(49) and (51) shows also the difference between the VD,
$\sigma^{v}_j$, and PVD, $\sigma^{pv}_j$. The VD contains the contributions
of $P(n)$ on all scales $n < 2^j$. That is, even for a small scale $L/2^j$,
$\sigma^{v}_j$ is mainly determined  by density perturbation on scales
larger tha $L/2^j$. The PVD, $\sigma^{pv}_j$, however is determined by
$P(n)$ on the scales $L/2^j$. From the DWT analysis, we have
$|\hat{\psi}(n)|^2=|\hat{\phi}(n/2)|^2-|\hat{\phi}(n)|^2$.
Therefore, the PVD is determined by the power of density perturbations in
the scale range from $L/2^j$ to $L/2^{j+1}$.

Generally, density perturbations on large scale is linear, while
non-linear on small scales. Therefore, eq.(49) would be a good
approximation for the VD, even on small scales, while eq.(51)
would be a poor approximation for the PVD on scales for which the
non-linear clustering is onset. This can be seen from Figs. 8 and
9. For the VD (Fig. 8), the theory [eq.(49)] and numerical
simulation gives basically coincident results. However, for the
PVD (Fig. 9), the result given by eq.(51) is significantly lower
than numerical simulation. Therefore, one may conclude that the
VD and PVD describes, respectively, the linear (large scales) and
non-linear (small scales) behavior of the cosmic mass field.

\subsection{One-point distribution of $v_{i}$ and $\Delta v_i$}

The PDF of the galaxy pair velocity is usually modeled as an exponential,
$f(\Delta v) \propto e^{-2^{1/2}\Delta v/\sigma^{pv}}$. This model
is supported by best fitting of early galaxy redshift surveys, such as
14.5 $m_b$ CfA (Davis \& Peebles 1993) and 1.2 Jy IRAS (Fisher et al.
1994.) Some non-linear clustering models also yield exponential
distribution of pair velocity (Sheth 1996, Diaferio \& Geller, 1996).
However, these results were measured by conventional techniques,
it cannot directly be applied to variables $v_{i \ {\bf j,l}}$ or
$\Delta v_{i \ {\bf j,l}}$. We calculate one-point distributions of
field variables $v_{i}$ and $\Delta v_{i}$.

For a given scale $j$, the 2$^{(j_1+j_2+j_3)}$ values of
$v_{i\ {\bf j,l}}$ (or
$\epsilon^{v_i}_{\bf j,l}/\epsilon_{\bf j,l}$) form an ensembles
of $v_{i\ {\bf j,l}}$. Thus, the one-point distributions $f(v_i)$
on scale $j$ can be obtained directly from the distribution of
$v_{i\ {\bf j,l}}$.

Figure 10 gives the one-point distribution $f(v_i)$ on scales $j=2
- 7$ of the SCDM. Models of $\tau$CDM and $\Lambda$CDM give the
similar results.
Figure 10 shows that the distributions are scale-dependent. On
large scales $j=2,3$ the distribution is close to a Gaussian. On
small scales $j=6,7$ the distribution in the middle range
($|v_i|<450$ km/s) can still be fitted with a Gaussian function.
In the range $400 < v_i < 1200$ km/s, $f(v_i)$, however, follows a
straight line which implies that the distribution is exponential.
With increasing $v_i >1200$ km/s, $f(v_i)$ decays slower than a
straight line. It indicates that the tail of the one-point
distribution $f(v_i)$ extends further than an exponential
distribution. Therefore, only in the middle velocity range, we
can fit the distribution $f(v_i)$ by an exponential.

 From Fig. 8 and eqs.(48) and (49), one can expect that the
PDF of VD measured by a conventional Gaussian window should be
about the same as Fig. 10. As has been mentioned, the scaling
function of the DWT analysis is just a window function. It will
give the similar statistical result as a Gaussian window if the
orthogonality and completeness are not the key of the relevant
statistics.

Figure 11 gives the one-point distribution $f(\Delta v_i)$ on
scales $j=2 - 7$ of the SCDM. Models of $\tau$CDM and $\Lambda$CDM
also give the similar results. The one-point distributions
$f(\Delta v_i)$ are
also scale-dependent, but it is significantly different from
$f(v_i)$. On large scale $j=2,3$ the distribution $f(\Delta v_i)$
is approximately exponential, i.e. $\ln f(\Delta v_i)$ vs. $\Delta
v_i$ can roughly be approximated by a straight line. However, on
all scales $j \geq 5$, the distribution $f(\Delta v_i)$ is
nothing but typically lognormal, i.e. $\ln[ f(\Delta
v_i)/f(2^a\Delta v_i)]=a \ln [f(\Delta v_i)/f(2\Delta v_i)]$,
where $a$ is a real number. Lognormal random field is often
employed to model the non-linear clustering (Cole \& Jones,
1991). The mass field traced by the Ly$\alpha$ forests is also
lognormal (Bi \& Davidsen, 1997; Feng \& Fang 2000). Moreover,
lognormal random field is typically intermittent (e.g.
Zel'dovich, Ruzmaikin, \& Sokoloff 1990). Based on these
considerations, it is not unusual to find that the one-point
distribution of $\Delta v_i$ is lognormal.

We also calculate the one-point distribution of pairwise velocity
measured by conventional pair counting. The result is ploted in
Fig. 12, which are the same as all measurements by pair counting
method, i.e. the distributions basically are exponential. Although
we have $\langle v^2_{12}(r) \rangle^{1/2} \simeq \sqrt{3}
\sigma^{pv_i}_j$ on small scales, the PDF of $v_{12}(r)$ is very
different from $\Delta v_i$ in the same scale range.

\section{Correlations between velocity and density fields}

\subsection{Local velocity-density correlation}

In redshift distortion theory, VD, $\sigma^v$, and its distribution
generally are assumed to be independent of the density field.
Observational evidences, however, seem
to support the existence of correlation between the velocity and
local number density of galaxies. For instance, the pairwise
velocity dispersion measured by the conventional method is found to
be sensitive to the presence of dense objects, like rich clusters
(e.g. Mo, Jing \& B\"{o}rner 1993). However, the conventional
method of measuring the peculiar or pairwise velocity is biased
to dense objects, due to the number-counting and
pair-counting-weighted statistic. Thus, the $v({\bf x})-n({\bf
x})$ correlation might be contaminated by the density-density
$n({\bf x})-n({\bf x})$ correlation. With the field variables
$v_{i \ {\bf j,l}}$ and $\Delta v_{i \ {\bf j,l}}$, one can
distinguish among the correlations of $v({\bf x})-n({\bf x})$
and $n({\bf x})-n({\bf x})$. Thus, one can detect the correlation
between the velocity and local density without bias.

Let us consider the number-counting statistics of
the mean peculiar velocity. It is given by
\begin{equation}
v_{i \ trad}=\frac{\sum_{n=1}^{N_g}w_nv_i({\bf x_n})}
  {\sum_{n=1}^{N_g} w_n}
\end{equation}
Using the ``partition of unity" of wavelet (Daubechies 1992), i.e.
$\sum_{{\bf l}=0}^{2^{\bf j}-1}\phi_{\bf j,l}^P({\bf x})
=(L_1L_2L_3)/2^{j_1+j_2+j_3}$, one can rewrite eq.(52) as
\begin{equation}
v_{i \ trad}=
\frac{\sum_{{\bf l}=0}^{2^{\bf j}-1}
  \sum_{n=1}^{N_g}w_nv({\bf x_n})\phi_{\bf j,l}({\bf x_n})}
{\sum_{{\bf l}=0}^{2^{\bf j}-1}
   \sum_{n=1}^{N_g} w_n \phi_{\bf j,l}({\bf x_n})} =
\frac{\sum_{{\bf l}=0}^{2^{\bf j}-1}\epsilon^{v_i}_{\bf j,l} }
  {\sum_{{\bf l}=0}^{2^{\bf j}-1}\epsilon^n_{\bf j,l}}.
\end{equation}

Therefore, the existence of the local $v$-$n$ correlation
can be tested by
\begin{equation}
C^{(v^2,n^2)}_{j \ trad}= \frac{2^{\bf j}\sum_{{\bf l}=0}^{2^{\bf
j}-1}
  [\epsilon^{v_i}_{\bf j,l}\epsilon^{n}_{\bf j,l}]^2}
{\sum_{{\bf l}=0}^{2^{\bf j}-1}|\epsilon^{v_i}_{\bf j,l}|^2
\sum_{{\bf l}=0}^{2^{\bf j}-1}|\epsilon^{n}_{\bf j,l}|^2},
\end{equation}
where we still consider the diagonal modes $j_1=j_2=j_3=j$. If there is
no correlation between the local velocity variable
$\epsilon^{v_i}_{\bf j,l}$ and density variable $\epsilon^{n}_{\bf j,l}$,
we have $C^{(v^2,n^2)}_{j \ trad}=1$. We display the result in
Fig. 13, which shows that for the CDM model
$C^{(v^2,n^2)}_{j \ trad} \gg 1$ on $k >1$ h Mpc$^{-1}$.
Nevertheless, we should not immediately conclude the existence of strong
correlation between $v({\bf x})$ and $n({\bf x})$, because Fig. 13 also
shows a strong deviation from $C^{(v^2,n^2)}_{j \ trad}=1$ for random2,
i.e. the samples without velocity-density correlation. That is,
the $C^{(v^2,n^2)}_{j \ trad} \gg 1$ given by random2 is completely from
density-density correlation.

To identify the $v$-$n$ correlation, it is essentially to refer to both
random1 and random2. The values of $C^{(v^2,n^2)}_{j \ trad}$ for the
CDM mode deviate not only from random1, but also random2. Thus, we may
conclude that the peculiar velocity given by the number-counting
statistics is correlated with density field, but not very strong.

For the DWT velocity variables $v_{i \ {\bf j,l}}$ [eqs.(19) and
(36)], the mean peculiar velocity is $(1/2^{\bf j})\sum_{{\bf
l}=0}^{2^{\bf j}-1}\epsilon^{v_i}_{\bf j,l}/\epsilon^n_{\bf j,l}$.
Thus, the local $v-n$ correlation can be measured by
\begin{equation}
C^{(v^2,n^2)}_j= \frac{2^{\bf j}\sum_{{\bf l}=0}^{2^{\bf j}-1}
  [v_{i \ {\bf j,l}}\epsilon^{n}_{\bf j,l}]^2}
{\sum_{{\bf l}=0}^{2^{\bf j}-1}|v_{i \ {\bf j,l}}|^2 \sum_{{\bf
l}=0}^{2^{\bf j}-1}|\epsilon^{n}_{\bf j,l}|^2} =\left
(\frac{\bar{\sigma}_j^v}{{\sigma}_j^v} \right )^2.
\end{equation}
That is, the ratio between $\bar{\sigma}_j^v$ and $\sigma_j^v$ is
a measure of the VD - $n$ correlation. The CDM spectra of
$\sigma_j^v$ and $\bar{\sigma}_j^v$ shown in Figs. 3 and 4 are
very similar. Therefore, in term of the DWT variables, the VD-$n$
correlation is very weak (see Fig. 13).

The traditional estimation of the velocity dispersion within
the patches $({\bf j,l})$ is given by
\begin{equation}
|\Delta v_{i \ trad}|=
\frac{\sum_{{\bf l}=0}^{2^{\bf j}-1}|\tilde{\epsilon}^{v_i}_{\bf j,l}| }
  {\sum_{{\bf l}=0}^{2^{\bf j}-1}\epsilon^n_{\bf j,l}}.
\end{equation}
It actually is the mean value of the absolute pairwise velocity
of models $({\bf j,l})$. The local correlation between
this pairwise velocity and density field can be estimated by
\begin{equation}
C^{(pv^2,n^2)}_{j \ trad}= \frac{2^{\bf j}\sum_{{\bf
l}=0}^{2^{\bf j}-1}
  [\tilde{\epsilon}^{v_i}_{\bf j,l}\epsilon^{n}_{\bf j,l}]^2}
{\sum_{{\bf l}=0}^{2^{\bf j}-1}|\tilde{\epsilon}^{v_i}_{\bf
j,l}|^2 \sum_{{\bf l}=0}^{2^{\bf j}-1}|\epsilon^{n}_{\bf j,l}|^2}.
\end{equation}
The result of $C^{(pv^2,n^2)}_{j \ trad}$ is given in Fig. 14.
Although for the CDM model, the correlation $C^{(pv^2,n^2)}_{j \ trad}$
is much larger than 1 on small scales, but it is always less than
the $C^{(pv^2,n^2)}_{j \ trad}$ of random2. Therefore, it seems to be
not an evidence of the correlation between pairwise velocity and local
density.

Using the field-variable-defined pairwise velocity
$\Delta v_{i \ {\bf j,l}}$, the correlation should be can be
measured by
\begin{equation}
C^{(pv^2,n^2)}_j= \frac{2^{\bf j}\sum_{{\bf l}=0}^{2^{\bf j}-1}
  [\Delta v_{i \ {\bf j,l}}\epsilon^{n}_{\bf j,l}]^2}
{\sum_{{\bf l}=0}^{2^{\bf j}-1}|\Delta v_{i \ {\bf j,l}}|^2
\sum_{{\bf l}=0}^{2^{\bf j}-1}|\epsilon^{n}_{\bf j,l}|^2} =\left
(\frac{\bar{\sigma}_j^{pv}}{{\sigma}_j^{pv}} \right )^2.
\end{equation}
Fig 14 also displays this ratio. For the CDM models, $C^{(pv^2,n^2)}_j$
deviates from $C^{(pv^2,n^2)}_j=1$ (random1), but it is
similar to that of random2. Therefore, no strong correlation
between the pairwise velocity and local density can be identified.

\subsection{Correlation between PVD and local density fluctuations}

To account for the redshift distortion effect in measuring the
DWT power spectrum, it would be necessary to investigate the
correlation between the density fluctuations $\delta({\bf x})$
and velocity field ${\bf v}({\bf x})$. The DWT power spectrum is
determined by the density fluctuations on mode $({\bf j,l})$. The
bulk velocity of mode $({\bf j,l})$ causes only a position shift
of mode $({\bf j,l})$ in redshift space with respect to real
space, but do not change the corresponding power of the density
fluctuations of mode $({\bf j,l})$. On the other hand, the
pairwise velocity of mode $({\bf j,l})$ will change the power of
the density fluctuations of mode $({\bf j,l})$. Therefore, the
redshift distortion of the DWT power spectrum is mainly caused by
velocity field variables $\Delta {\bf v}_{\bf j,l}$, not
${\bf v}_{\bf j,l}$. We will only study the correlation between
$\Delta {\bf v}_{\bf j,l}$ and density fluctuation
$\delta({\bf x})$.

A simplest statistical measure of the
$\Delta {\bf v}_{\bf j,l}$-$\delta$ correlation is the second
order statistic defined as
\begin{equation}
C^{(pv, \delta)}_j= \frac{2^{\bf j}\sum_{{\bf l}=0}^{2^{\bf
j}-1}\tilde{\epsilon}^{v^i}_{\bf j,l}
  \tilde{\epsilon}^{n}_{\bf j,l}}
{\sum_{{\bf l}=0}^{2^{\bf j}-1}|\tilde{\epsilon}^{v^i}_{\bf j,l}|
\sum_{{\bf l}=0}^{2^{\bf j}-1}|\tilde{\epsilon}^{n}_{\bf j,l}|},
\end{equation}
$C^{(pv,\delta)}$ will be equal to zero, if there is no
correlation between velocity fluctuation WFC
$\tilde\epsilon^{v_i}_{\bf j,l}$ and density fluctuation WFC
$\tilde\epsilon^n_{\bf j,l}$.

Fig. 15 shows $C^{(pv, \delta)}_j$ in a CDM model, all
results are consistent with $C^{(pv, \delta)}_j=0$ within the
error bars. Thereby, in terms of second order statistics, it is
likely reasonable assumption that there is no $\Delta {\bf v}_{\bf
j,l}$-$\delta$ correlation. However, $C^{(pv,\delta)}_j=0$ doesn't
mean that $\Delta {\bf v}_{\bf j,l}$ and $\delta({\bf x})$ are
uncorrelated in general. Since the one-point distribution of
$\Delta {\bf v}_{\bf j,l}$ are invariant with respect to $\Delta
v_i({\bf x}) \rightleftharpoons - \Delta v_i({\bf x})$, in this
case, statistics with linear term of $\Delta v_i({\bf x})$,
like $\langle \Delta v_i({\bf x})\delta({\bf x})\rangle$ will be
zero. Therefore, we should study the correlation between $|\Delta
v_i({\bf x})|^2$ and $\delta^2({\bf x})$, which can be calculated
by
\begin{equation}
C^{(pv^2, \delta^2)}_j= \frac{2^{\bf j}\sum_{{\bf l}=0}^{2^{\bf
j}-1}[\tilde{\epsilon}^{v^i}_{\bf j,l} \tilde{\epsilon}^{n}_{\bf
j,l}]^2} {\sum_{{\bf l}=0}^{2^{\bf
j}-1}|\tilde{\epsilon}^{v^i}_{\bf j,l}|^2 \sum_{{\bf
l}=0}^{2^{\bf j}-1}|\tilde{\epsilon}^{n}_{\bf j,l}|^2}.
\end{equation}
The results for the dark matter model are also plotted in Fig. 15.
It shows that $C^{(pv^2, \delta^2)}_j$ of the CDM models is higher
than that of random1 and random2. Therefore, $\Delta {\bf v}$ and
$\delta({\bf x})$ are locally correlated on scales $k> 0.2 $ h
Mpc$^{-1}$.

\section{Conclusion}

It is known that in the DWT representation a continuous field, like
cosmic density field $n({\bf x})$, can be described by discrete
variables, given by the DWT scale-space decomposition of $n({\bf x})$
into the DWT mode  $({\bf j,l})$ (Fang \& Feng 2000). Since the modes
$({\bf j,l})$ are complete and orthogonal, the DWT decomposition
is information-lossless and not redundant. It insures that one can
compare statistical results of different samples.

Taking the advantage of the DWT analysis, we show that the cosmic
velocity field can also be properly described by the discrete
variables $v_{i \ {\bf j,l}}$, which are given by an assignment
of the number density and velocity of galaxies (or particles)
into the DWT modes $({\bf j,l})$. In this scheme, the peculiar
velocity and pairwise velocity of galaxies or particles are given
by field variables, which correspond to the coefficients of the
scaling functions and wavelets of the DWT decomposition. As a
consequence, the VD and PVD are no longer measured by
number-counting or pair-counting statistic, but with the ensemble
of the field variables.

Using simulation samples, we show that peculiar velocity field is
significantly different from randomized field. The peculiar
velocities show correlation on scales of few tens
h$^{-1}$ Mpc. The pairwise velocity (or relative velocity) has
similar correlation. On small scales, we also found significant
correlations between pairwise velocity and density fluctuations.
This is especially valuable in treating the redshift distortion
of the DWT power spectrum.

Although the VD and PVD look similar to each other in some
aspects, they actually are very different both statistically and
physically. The VD-to-PVD ratio shows the difference between the
scale-dependencies of the VD and PVD. More prominent difference
between the VD and PVD is shown by one-point distribution. The
one-point distribution of the VD is approximately exponential,
while the PVD is lognormal. This difference is typical of
intermittent field (a similar difference
between the one-point distributions of field variable $A$ and
the difference $\Delta A$ in turbulence is given by Shraiman
\& Sigglia 2000). Therefore, one can conclude that the cosmic
velocity field is intermittent.

\acknowledgments

LLF and YQC acknowledges support from the National Science
Foundation of China (NSFC) and National Key Basic Research Science
Foundation.

\appendix

\section{Poisson noise}

The observed galaxy number density distribution $n({\bf x})$
[eq.(2)] is believed to be a realization of a Poisson point
process with an intensity $\tilde{n}({\bf x})=\bar{n}^g({\bf
x})[1+\delta({\bf x})]$, where $\bar{n}^g({\bf x})$ is selection
function, and $\delta({\bf x})=[n({\bf x})/\bar{n}^g({\bf x})]-1$
is the density contrast fluctuations of the underlying matter
field. Thus, the correlation function of $\delta({\bf x})$ is
\begin{equation}
\langle \delta({\bf x})\delta({\bf x})\rangle = -1 +
\left \langle \frac{n({\bf x})n({\bf x'})}
{\bar{n}^g({\bf x})\bar{n}^g({\bf x'})}\right \rangle -
\delta_D({\bf x-x'})\frac{1}{\bar{n}^g({\bf x'})}.
\end{equation}
The power spectrum of $\delta({\bf x})$ is then given by
(Fang \& Feng 2000, Yang et al. 2001)
\begin{equation}
P_{\bf j}=I_{\bf j}^2-N_{\bf j}.
\end{equation}
where
\begin{equation}
I_{\bf j}^2 =\frac{1}{2^{(j_1+j_2+j_3)}}
\sum_{l_1=0}^{2^{j_1}-1}\sum_{l_2=0}^{2^{j_2}-1}\sum_{l_3=0}^{2^{j_3}-1}
\frac{[\tilde{\epsilon}^{n}_{\bf j,l}]^2}
  {[\bar{n}^{g}_{\bf j,l}]^2}
\end{equation}
and
\begin{equation}
N_{\bf j} =\frac{1}{2^{(j_1+j_2+j_3)}}
\sum_{l_1=0}^{2^{j_1}-1}\sum_{l_2=0}^{2^{j_2}-1}\sum_{l_3=0}^{2^{j_3}-1}
\frac{1}{\bar{n}^{g}_{\bf j,l}}.
\end{equation}
where the WFC $\tilde{\epsilon}^{g}_{\bf j,l}$ is given by
\begin{equation}
\tilde{\epsilon}^{n}_{\bf j,l}=
  \int n({\bf x}) \psi_{\bf j,l}({\bf x})d{\bf x}
\end{equation}
and $\bar{n}^{g}_{\bf j,l}$ is the mean number density of the selection
function $\bar{n}^g({\bf x})$ in the mode $({\bf j,l})$.

The term $I_{\bf j}^2$ is the mean power of ${\bf j}$ modes measured
from the observed realization $n^g({\bf x})$, and the term $N_{\bf j}$
is the power on ${\bf j}$ modes due to the Poisson noise. Therefore,
the Poisson error for the galaxy distribution $n({\bf x})$ can be
described by an error distribution as
\begin{eqnarray}
\lefteqn{ n^{error}({\bf x}) =
  \sum_{l_1=0}^{2^{J_1}-1} \sum_{l_2=0}^{2^{J_2}-1}
  \sum_{l_3=0}^{2^{J_3}-1}} \\ \nonumber
 & & \left [
  \frac{L_1L_2L_3}{2^{J_1+J_2+J_3}}\bar{n}^g_{\bf j,l} \right ]^{1/2}
\delta^D(x_1-l_1L_1/2^{J_1})\delta^D(x_2 -l_2L_2/2^{J_2})
\delta^D(x_3-l_3L_3/2^{J_3}).
\end{eqnarray}
where $(J_1,J_2,J_3)$ correspond to the smallest scale of sample
considered. In the case of $\bar{n}^g({\bf x})=$ const and
$L_1=L_2=L_3=L$, $J_1=J_2=J_3=J$,
Eq.(A6) becomes
\begin{equation}
n^{error}({\bf x}) =
 \sum_{{\bf l}=0}^{2^{\bf J}-1}\sqrt{\frac{N_g}{2^{3J}}}
  \delta^D({\bf x}- {\bf l}L/2^{J})
\end{equation}
It simply means that the Poisson noise of the galaxy number in a cell
$({\bf J,l})$ is $\sqrt{N_g/2^{3J}}$.

\section{Off-counting algorithm}

The typical terms in eqs.(41) and (43) are $\epsilon^{v_i}_{\bf
j,l}/\epsilon^{n}_{\bf j,l}$ and $\tilde\epsilon^{v_i}_{\bf
j,l}/\epsilon^{n}_{\bf j,l}$, where the denominator is given by
eq.(39)
\begin{equation}
\epsilon^{n}_{\bf j,l}=\int n({\bf x})\phi_{\bf j,l}({\bf x})d{\bf x}
    =\sum_{n=1}^{N_g}w_n \phi_{\bf j,l}({\bf x}_n).
\end{equation}
Therefore, for modes ${\bf j,l}$ corresponding to position of the
voids of galaxies, i.e. $n({\bf x})$ is very small or zero, the
SFC $\epsilon^{n}_{\bf j,l}$ would be very small, even zero.
These modes will lead to very large term of $\epsilon^{v_i}_{\bf
j,l}/\epsilon^{n}_{\bf j,l}$. Actually, these terms have very
large error, because the number density $n({\bf x})$ in these
modes generally is less than the error distribution at the same
spatial range $n^{error}({\bf x})$.

To avoids the contamination of these large error term, we use the
following algorithm.

1. Perform  DWT for the number density distributions of $n({\bf x})$, and
   $n^{error}$.

2. The SFC of $n({\bf x})$ generally is larger than the SFCs of
   $n^{error}({\bf x})$, and therefore
   $\epsilon^n_{\bf j,l} >\epsilon^{error}_{jl}$.
   However, for some modes $({\bf j,l})$, we have
\begin{equation}
\epsilon^n_{j,l} \leq \epsilon^{error}_{\bf j,l}
\end{equation}
These $({\bf j,l})$ are noise dominated modes, and no observed
data are available.

3. We eliminate all modes affected by the noise dominated modes.
This is, if $({\bf j,l})$ satisfies eq.(B2), we off-count the
following modes
\begin{eqnarray}
j'_i  & \geq & j_i \\ \nonumber
l'_i &  = & l2^{j'_i-j_i}....(l+1)2^{j'_i-j_i}-1.
\end{eqnarray}

This is called off-counting algorithm. It is effectively to
eliminate the effects of $\epsilon^{n}_{\bf j,l}\simeq 0$, and
give reasonable estimation of VD [eq.(41)] and PVD [eq.{43)] on
the same scale range as the power spectrum.

\newpage

\begin{figure}
\figurenum{1} \epsscale{0.6}
\plotfiddle{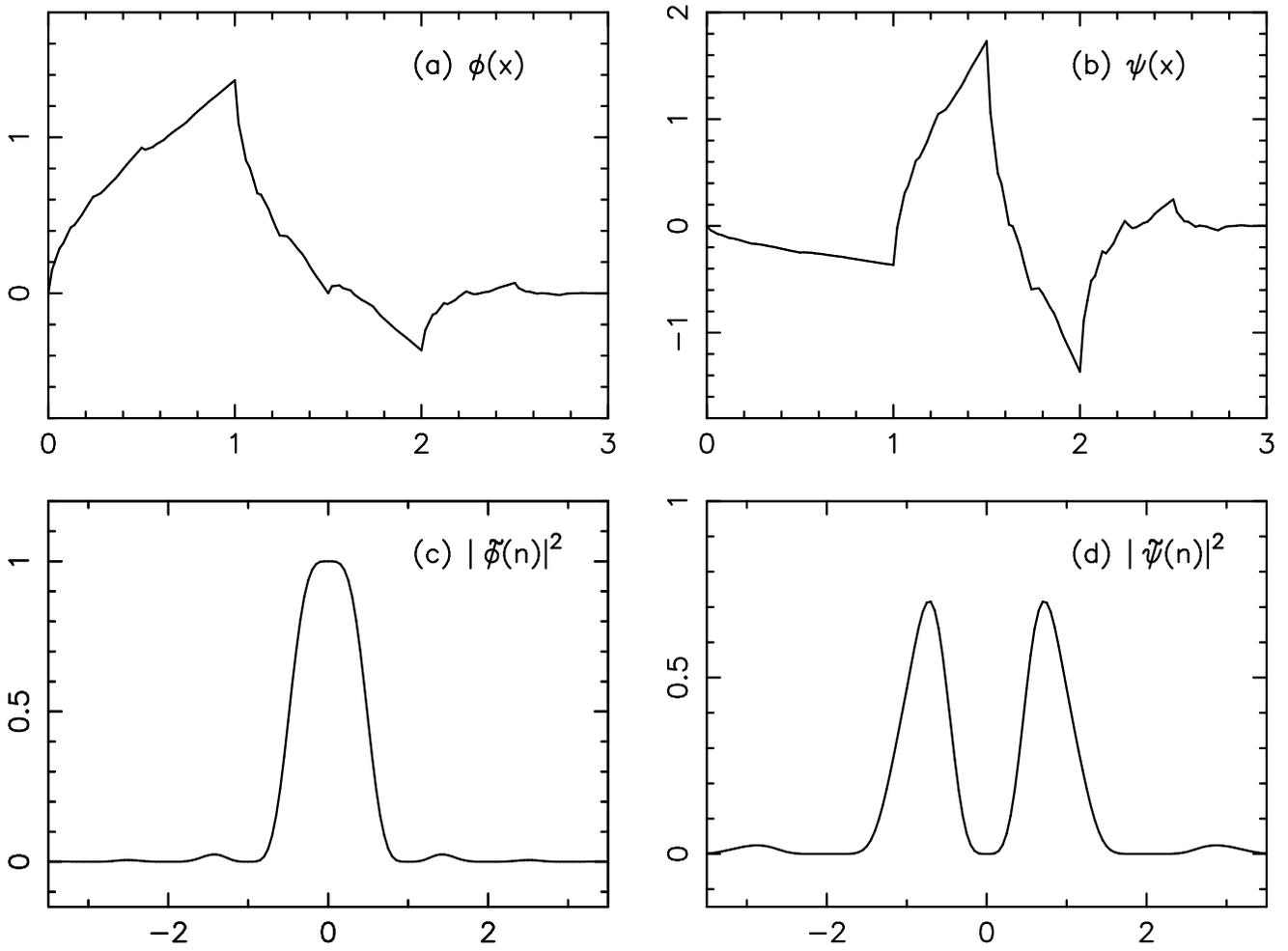}{20.cm}{270}{80}{80}{-310.}{470.}
\caption[]{The basic scaling function (a), basic wavelet (b) and
their Fourier transforms (c) and (d) of Daubechies 4 wavelet.}
\label{Fig1}
\end{figure}

\begin{figure}
\figurenum{2} \epsscale{0.6}
\plotfiddle{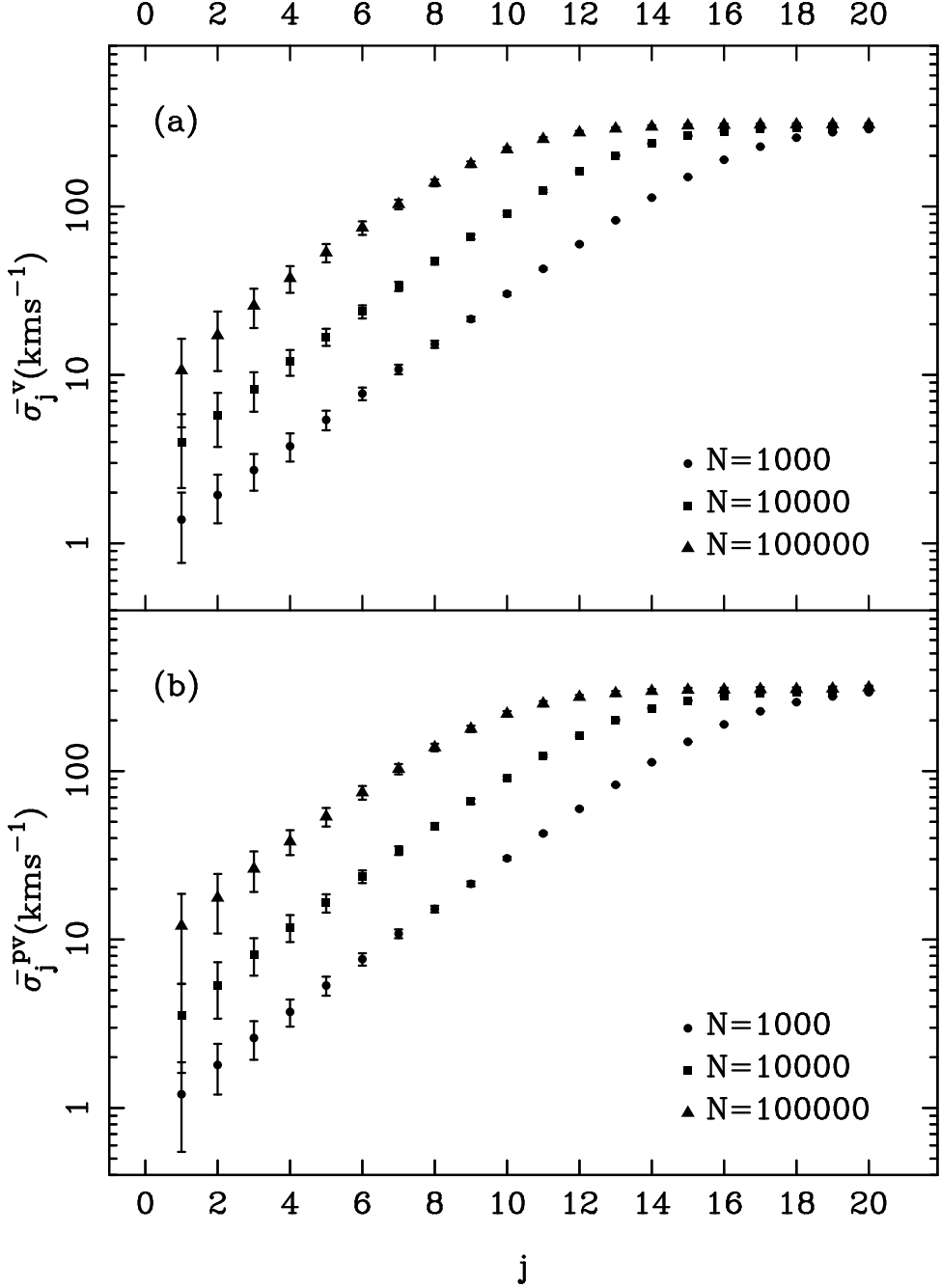}{20.cm}{0}{80}{80}{-230.}{-50.}
\caption[]{The VD spectrum, $\bar{\sigma}^v_j$ [panel (a)] and
the PVD spectrum, $\bar{\sigma}^{pv}_j$ [panel (b)] of 1-D random
samples, which are uniform random  distribution of $N$ particles
in spatial range $L=1$. Each particles assigned with a velocity
$v_n$ drawn from a Gaussian distribution with variance 300 km
s$^{-1}$. $j$ is for scale $1/2^j$. The error bars are 1-$\sigma$
variance from 1000 realizations} \label{Fig2}
\end{figure}

\begin{figure}
\figurenum{3} \epsscale{0.6}
\plotfiddle{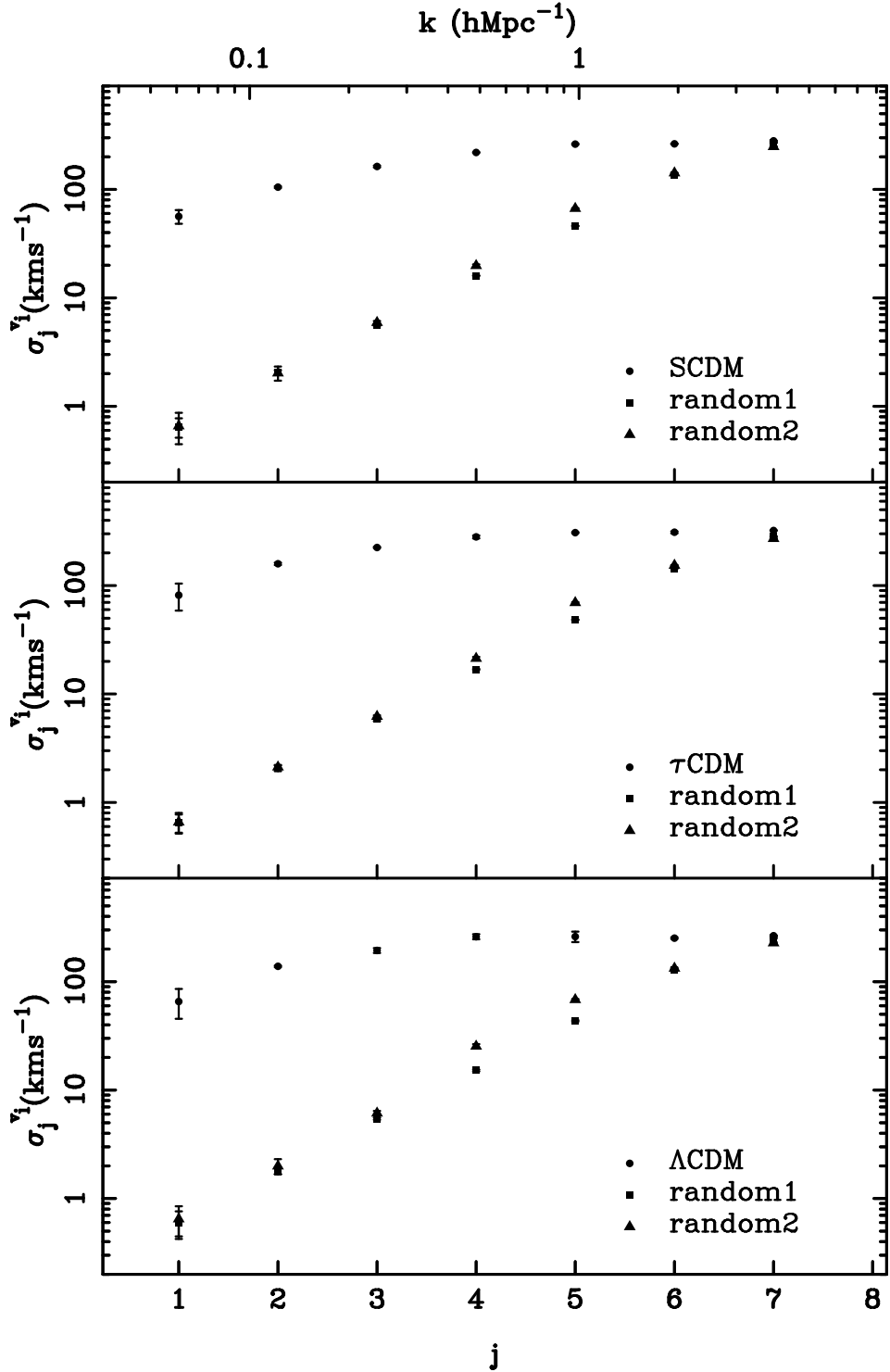}{20.cm}{0}{80}{80}{-230.}{-40.}
\caption[]{The VD spectrum of the diagonal mode, $\sigma^{v_i}_j$
of simulation samples for models SCDM (upper panel), $\tau$CDM
(central panel) and $\Lambda$CDM (lower panel). The random1
samples are given by assigning $N$ velocities to randomly
distributed particles, and random2 is produced by randomizing the
assignment of $N$ velocities to $N$ mass particles. The error bars
are 1-$\sigma$ variance from 10 realizations for each model.}
\label{Fig3}
\end{figure}

\begin{figure}
\figurenum{4} \epsscale{0.6}
\plotfiddle{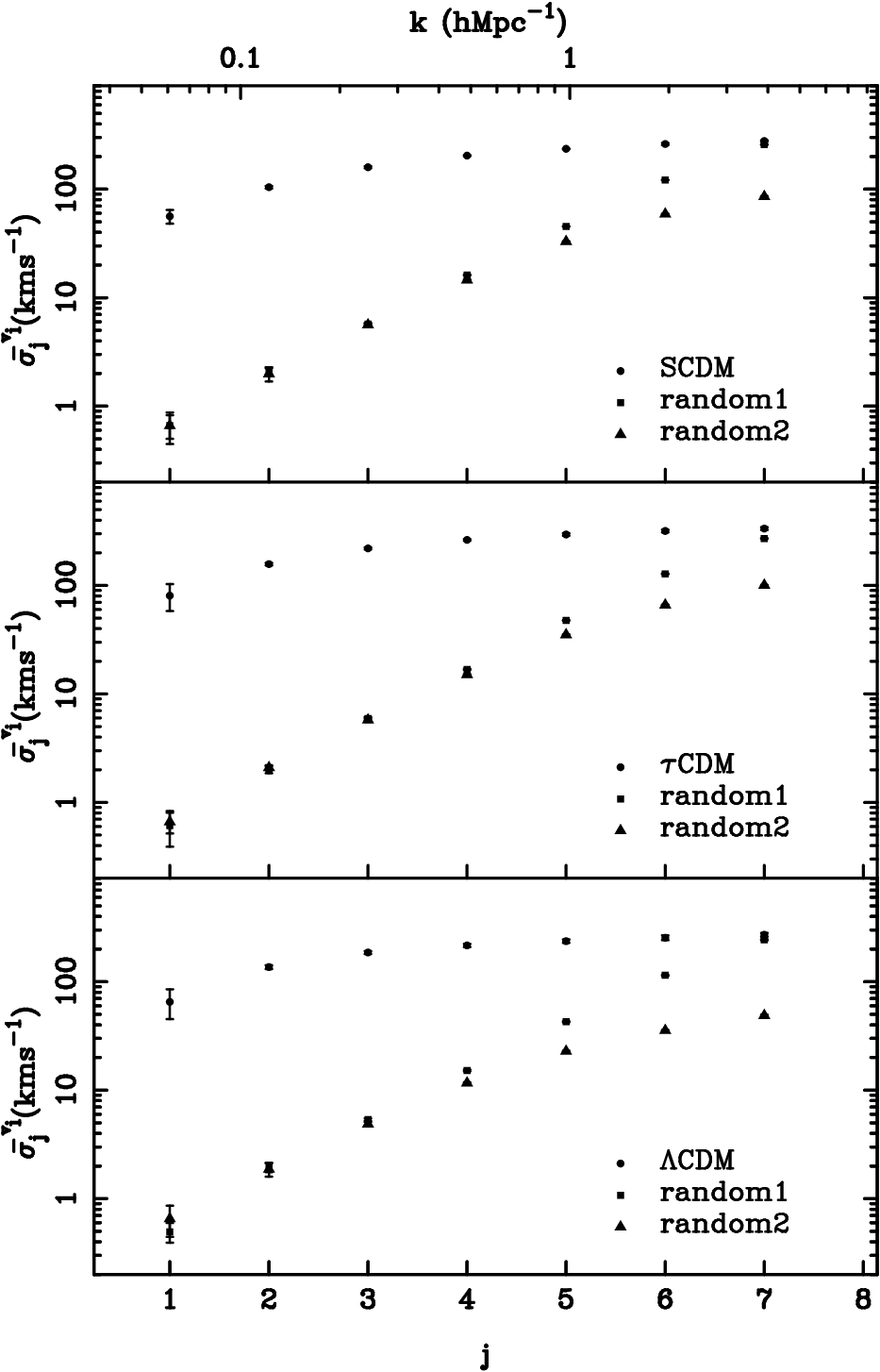}{20.cm}{0}{80}{80}{-230.}{-50.}
\caption[]{The  modified VD spectrum, $\bar{\sigma}^{v_i}_j$ of
diagonal mode $(j_1=j_2=j_3=j)$ for SCDM simulation samples
(upper panel), $\tau$CDM (central panel) and $\Lambda$CDM (lower
panel). The random samples are given by the randomizations as
Fig.3. The error bars are 1-$\sigma$ variance from 10 realizations
for each model. } \label{Fig24}
\end{figure}

\begin{figure}
\figurenum{5} \epsscale{0.6}
\plotfiddle{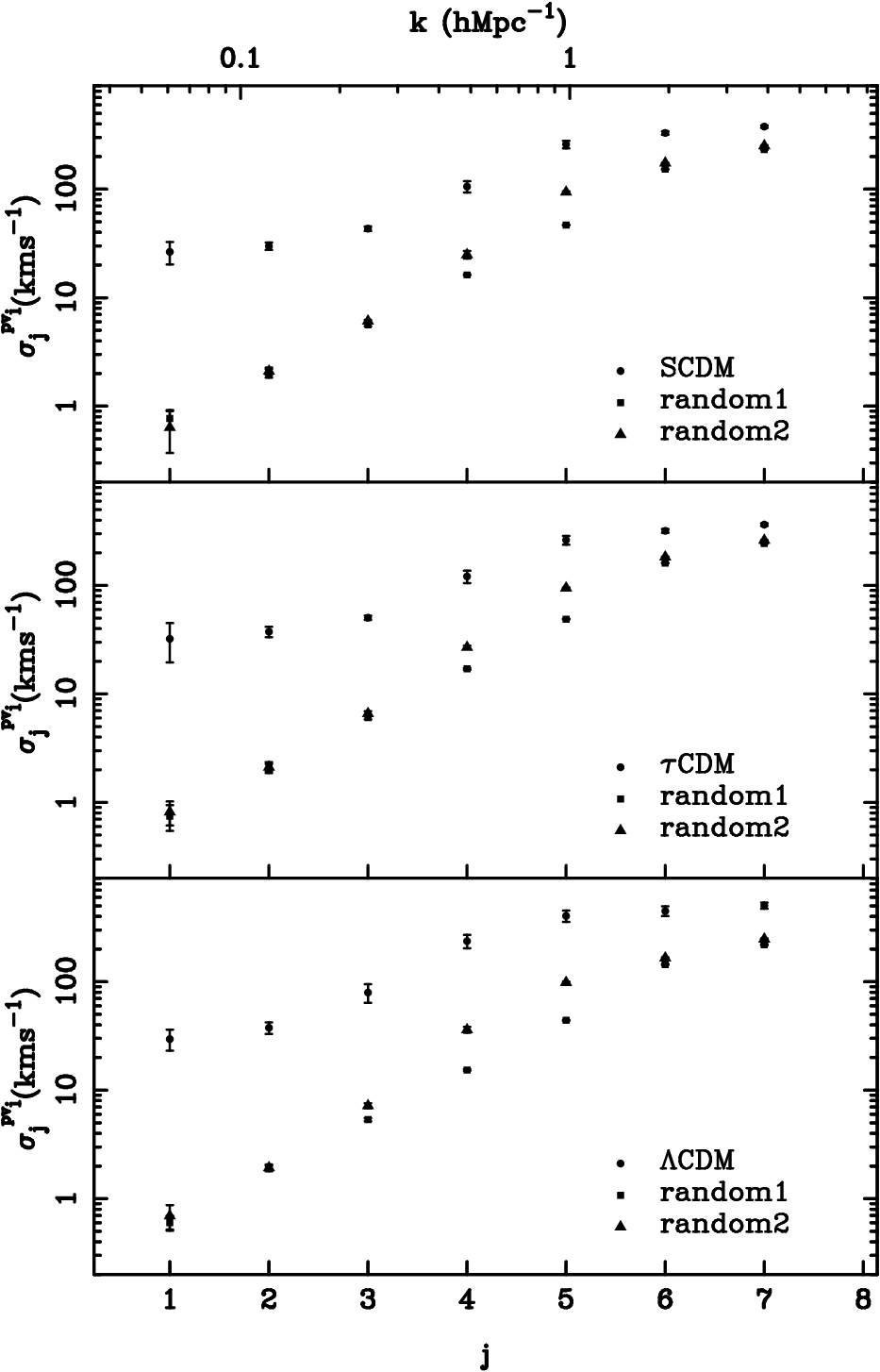}{20.cm}{0}{80}{80}{-230.}{-50.}
\caption[]{The PVD spectrum, $\sigma^{pv_i}_j$, of diagonal mode
$(j_1=j_2=j_3=j)$ for SCDM simulation samples (upper panel),
$\tau$CDM (central panel) and $\Lambda$CDM (lower panel). The
random samples are given by the randomizations as Fig. 3. The
error bars are 1-$\sigma$ variance from 10 realizations for each
model.} \label{Fig5}
\end{figure}

\begin{figure}
\figurenum{6} \epsscale{0.6}
\plotfiddle{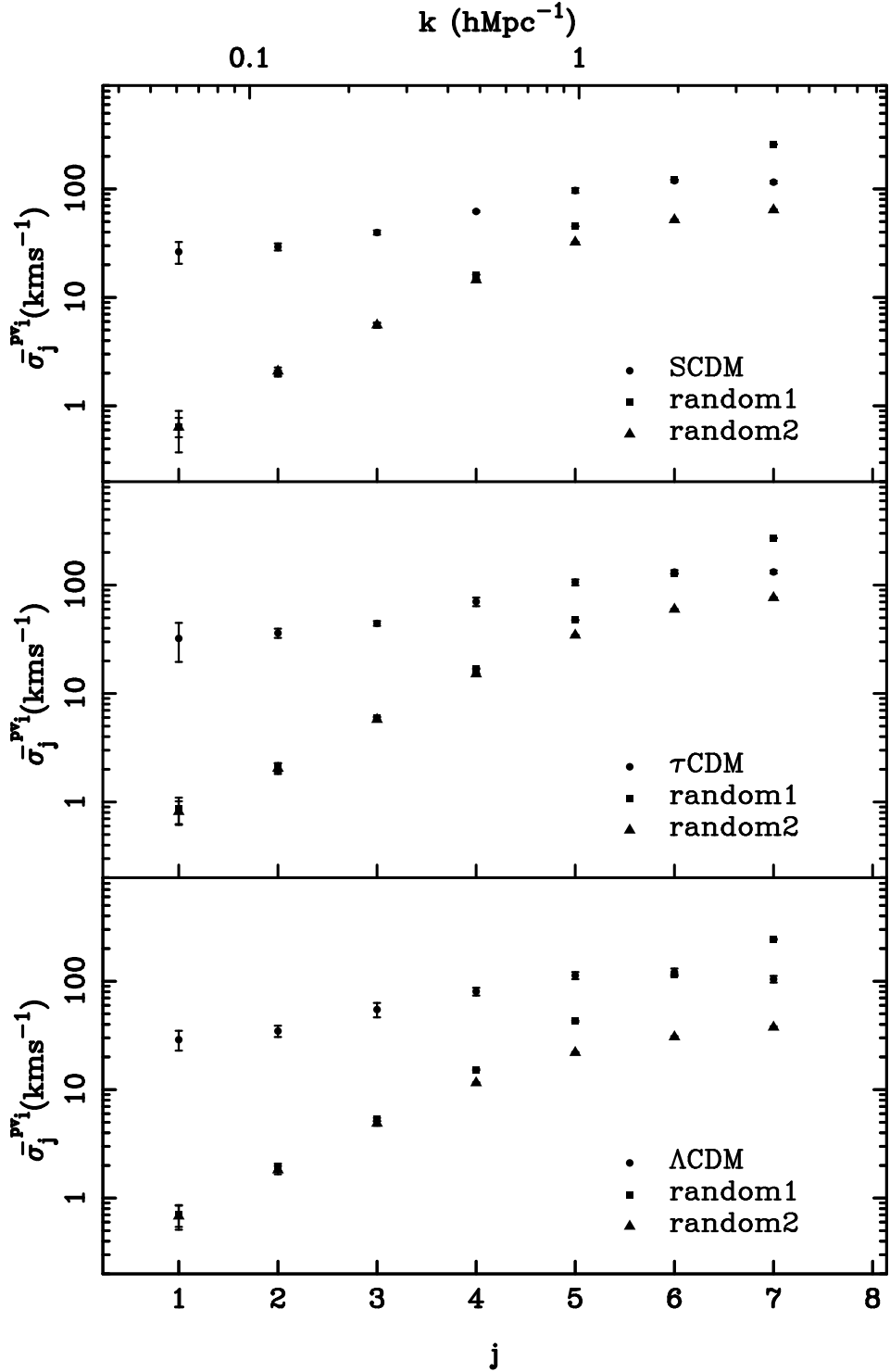}{20.cm}{0}{80}{80}{-230.}{-50.}
\caption[]{The modified PVD spectrum, $\bar{\sigma}^{pv_i}_j$, of
diagonal mode $(j_1=j_2=j_3=j)$ for SCDM simulation samples
(upper panel), $\tau$CDM (central panel) and $\Lambda$CDM (lower
panel). The random samples are given by the randomizations as Fig.
3. The error bars are 1-$\sigma$ variance from 10 realizations
for each model.} \label{Fig6}
\end{figure}

\begin{figure}
\figurenum{7} \epsscale{0.6}
\plotfiddle{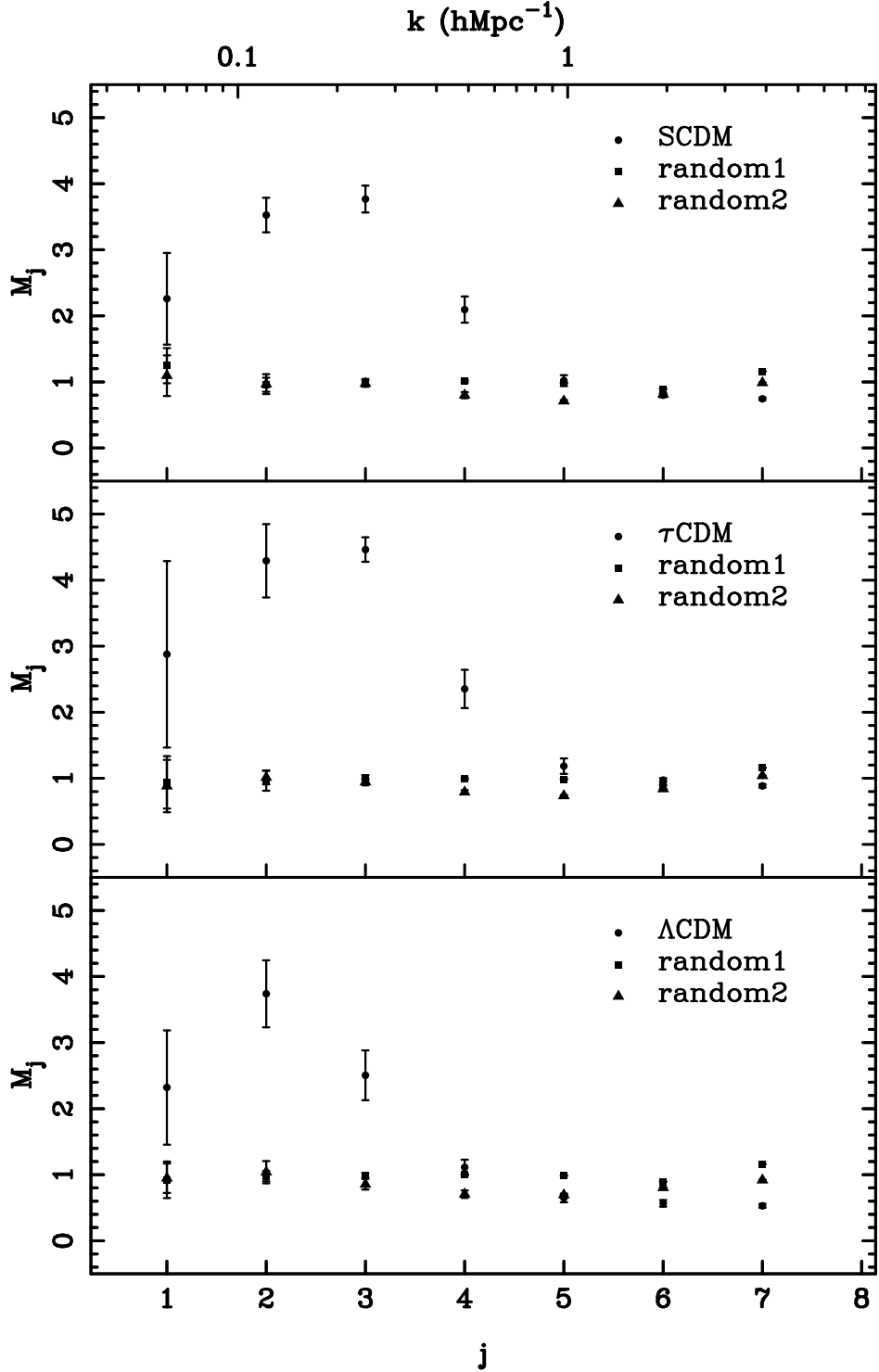}{20.cm}{0}{80}{80}{-230.}{-50.}
\caption[]{The VD-to-PVD ratio or Mach number $M_j$, of
simulation samples for models  SCDM  (upper panel), $\tau$CDM
(central panel) and $\Lambda$CDM (lower panel). The random samples
are the same as Fig. 3. The error bars are 1-$\sigma$ variance
from 10 realizations for each model.} \label{Fig7}
\end{figure}

\begin{figure}
\figurenum{8} \epsscale{0.6}
\plotfiddle{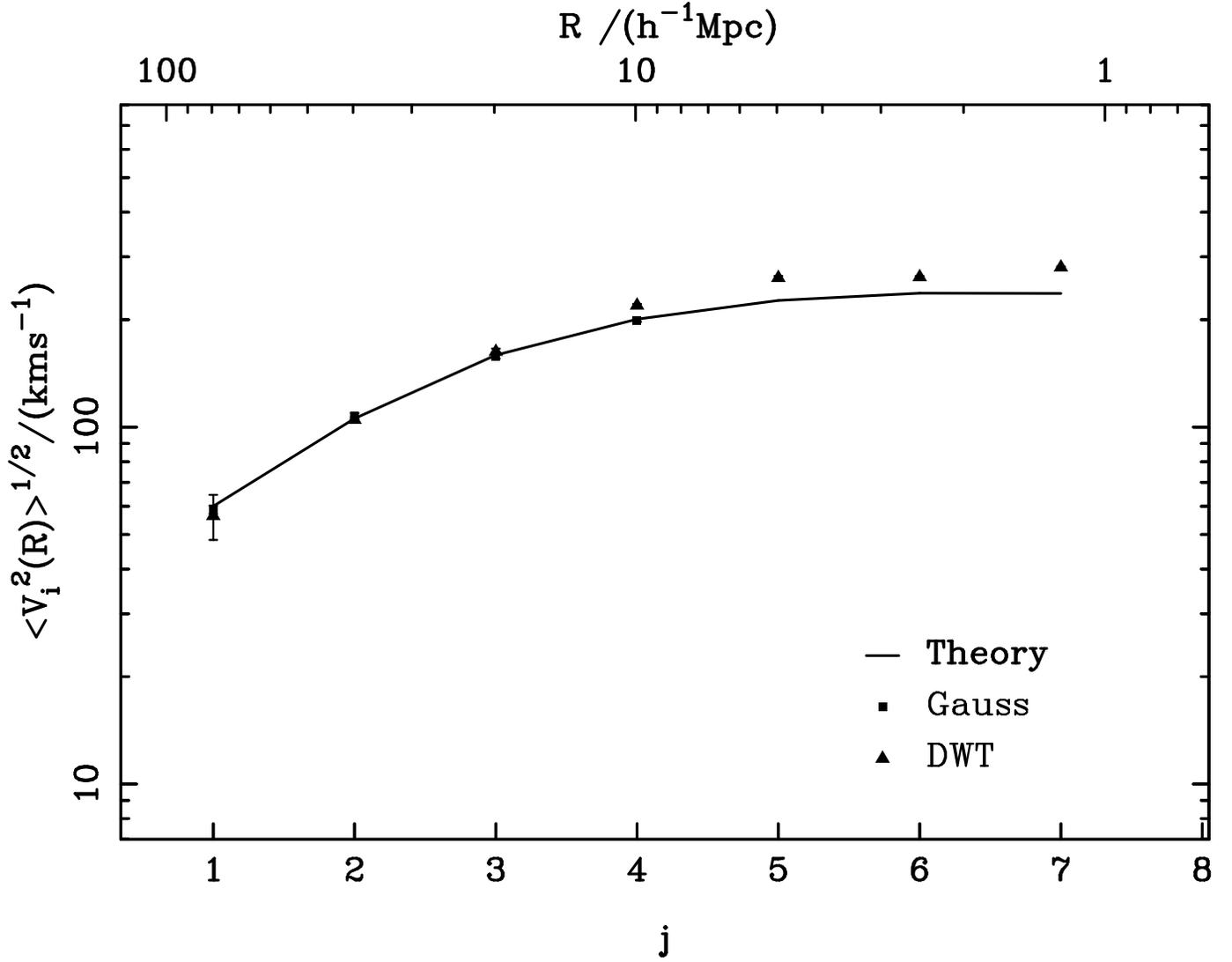}{20.cm}{270}{80}{80}{-310.}{470.}
\caption[]{A comparison between the velocity dispersion $\langle
V^2_i(R)\rangle^{1/2}$ measured by a Gaussian window function
(square) and by the DWT method (triangle). The theory value
(line) is the expected linear regime velocity using eq. (49). $R$
is the size of the window.} \label{Fig8}
\end{figure}

\begin{figure}
\figurenum{9} \epsscale{0.6}
\plotfiddle{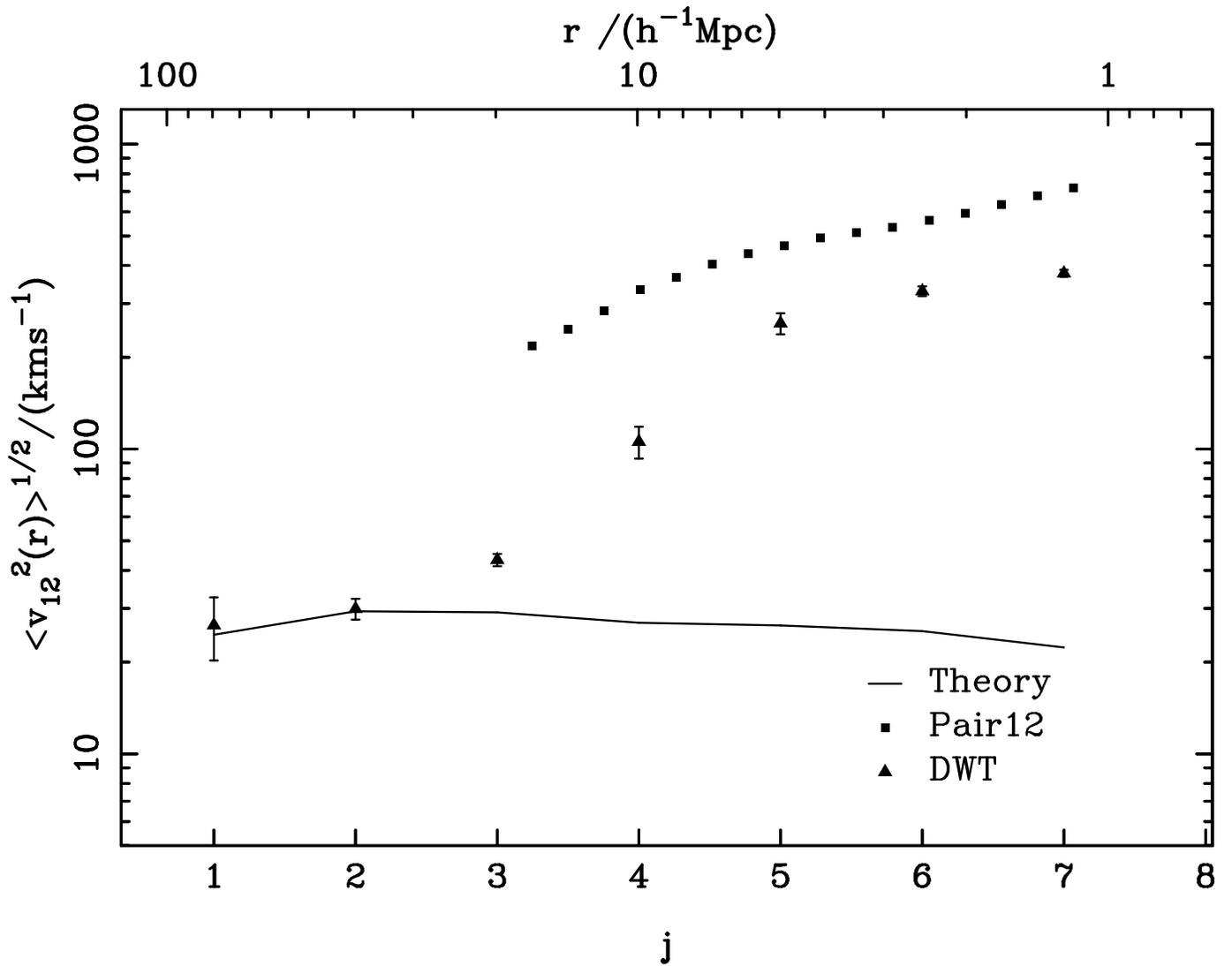}{20.cm}{270}{80}{80}{-310.}{470.}
\caption[]{A comparison between the pair velocity dispersion
$\langle v^2_{12}(r)\rangle^{1/2}$ measured by a traditional pair
counting method (square) and by the DWT method (triangle). $r$ is
the distance of the pairs. The theory value (line) is calculated
by eq. (51).} \label{Fig9}
\end{figure}

\begin{figure}
\figurenum{10} \epsscale{0.6}
\plotfiddle{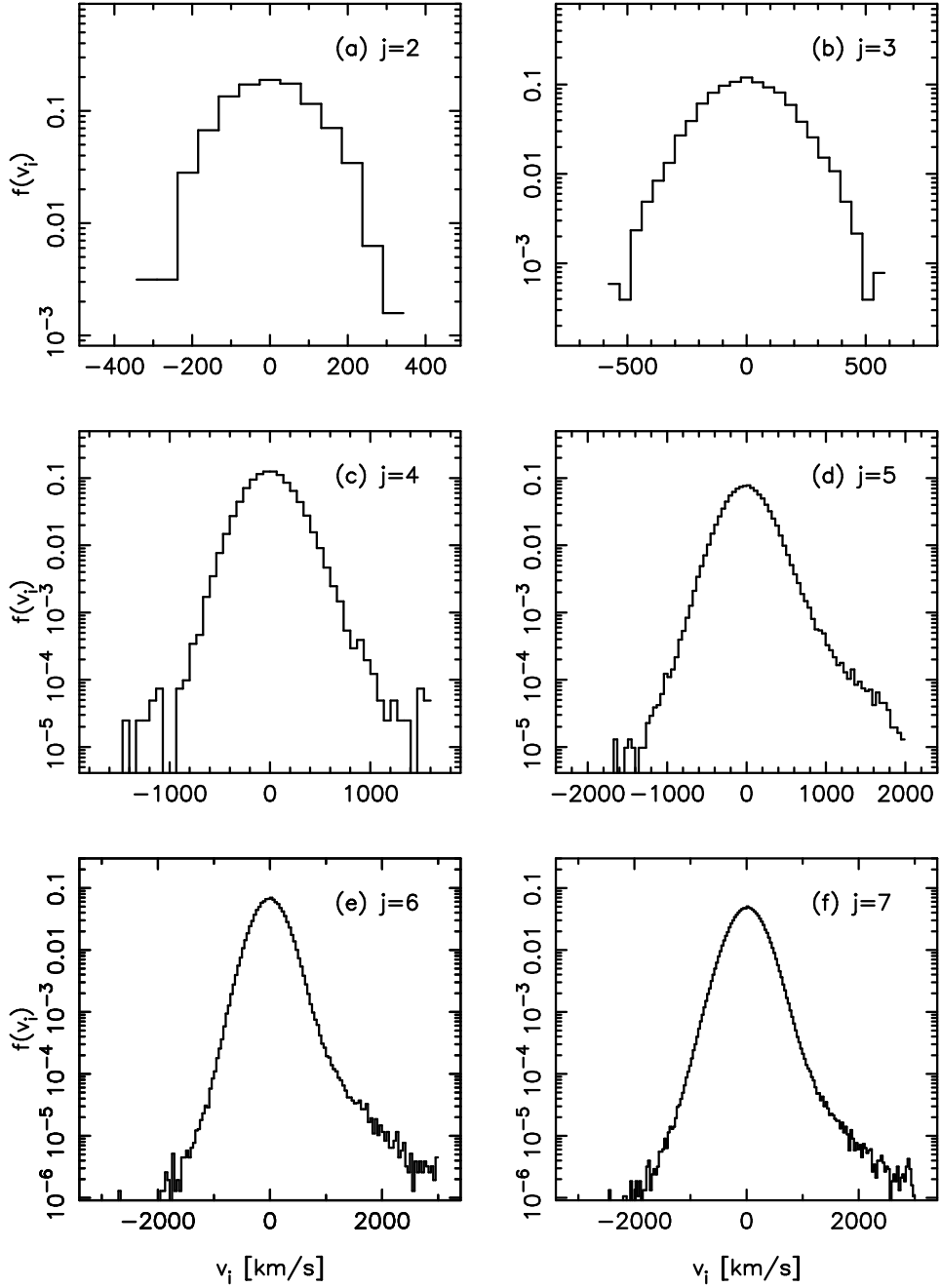}{20.cm}{0}{80}{80}{-230.}{-50.}
\caption[]{The one-point distribution of $v_i$ on scales $j=2 -
7$ for simulation samples of the SCDM model.} \label{Fig10}
\end{figure}

\begin{figure}
\figurenum{11} \epsscale{0.6}
\plotfiddle{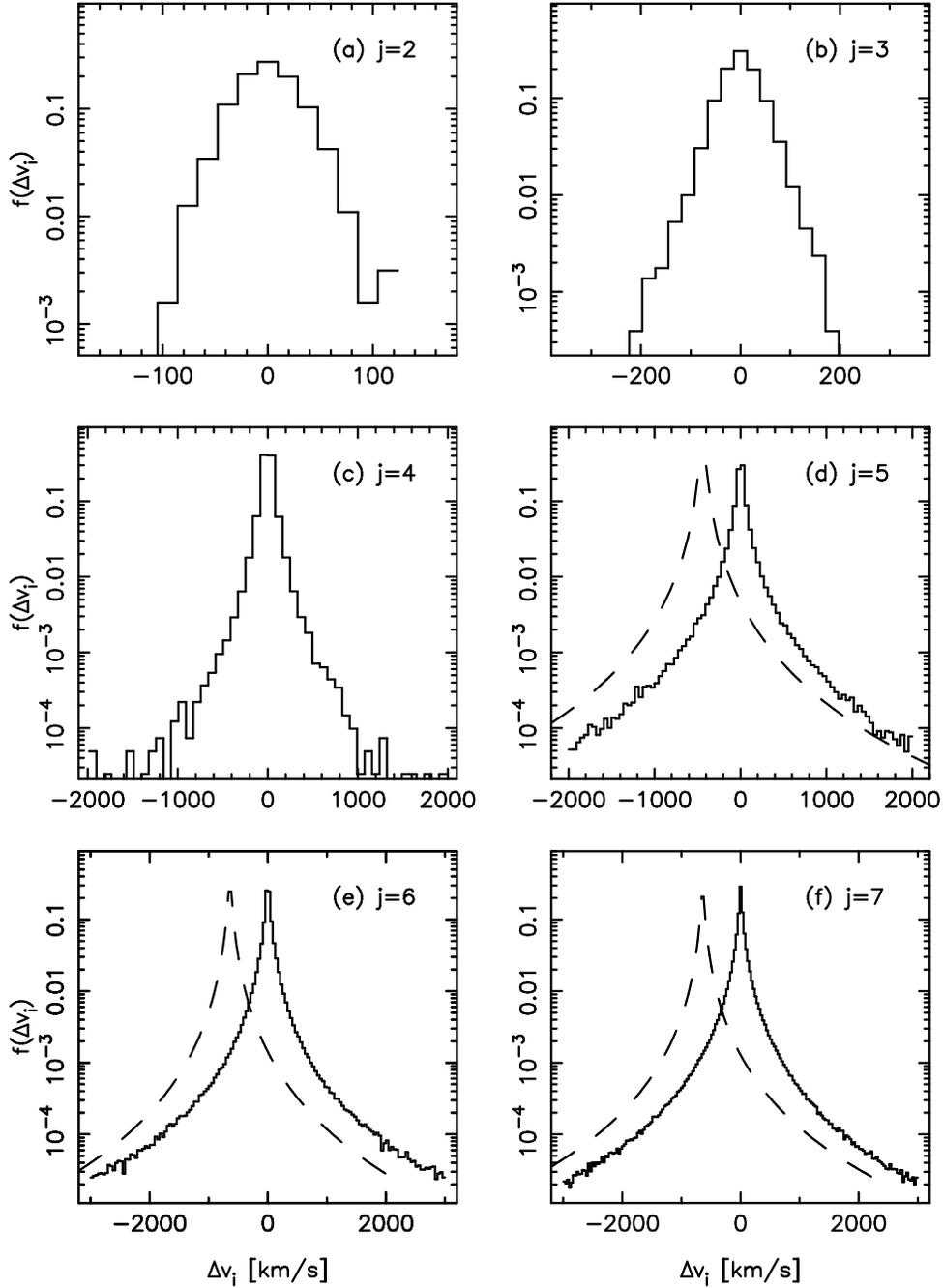}{20.cm}{0}{80}{80}{-230.}{-50.}
\caption[]{The one-point distribution of $\Delta v_i$ on scales
$j=2 - 7$ for simulation samples of the SCDM model. The dashed
line for j=5,6,7 is the lognormal curve with the same dispersion
as the CDM model, which are slightly shifted to left for a clear
presentation.} \label{Fig11}
\end{figure}

\begin{figure}
\figurenum{12} \epsscale{0.6}
\plotfiddle{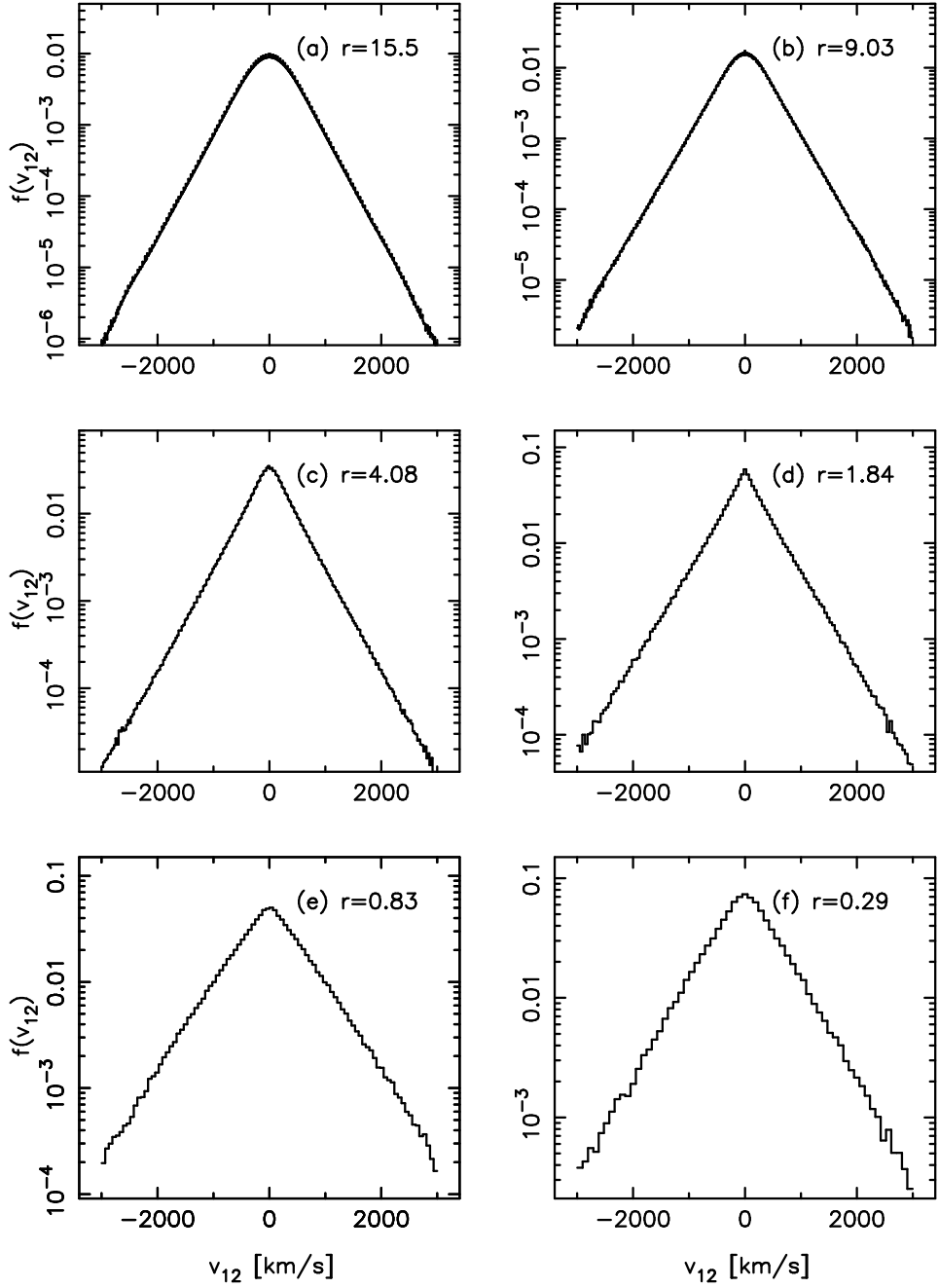}{20.cm}{0}{80}{80}{-230.}{-50.}
\caption[]{The one-point distribution of $v_{12}$ measured by the
traditional pair counting method for simulation samples of the
SCDM model. $r$ is the distance of the pairs in the unit of
h$^{-1}$ Mpc.} \label{Fig12}
\end{figure}

\begin{figure}
\figurenum{13} \epsscale{0.6}
\plotfiddle{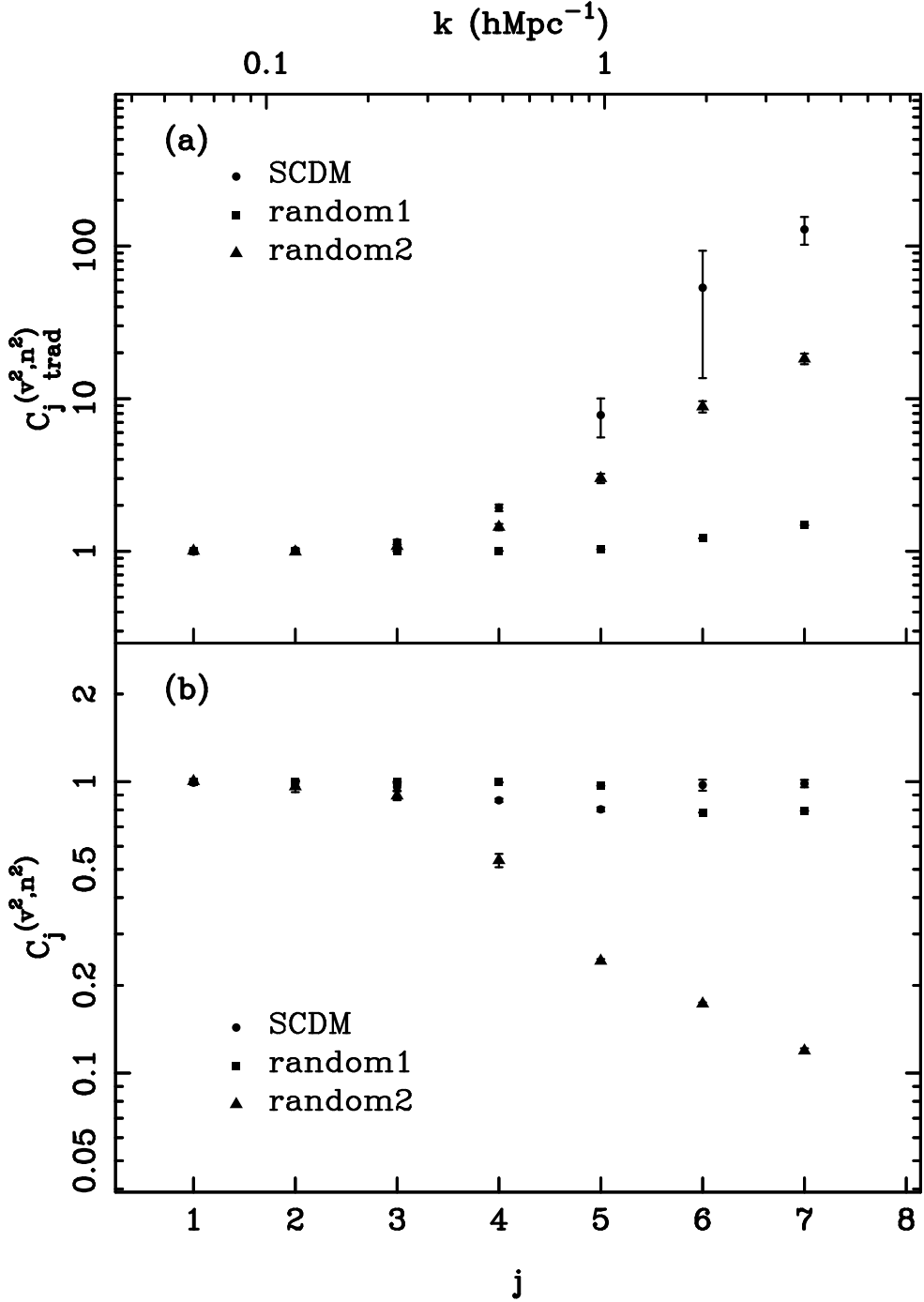}{20.cm}{0}{80}{80}{-230.}{-50.}
\caption[]{The fourth order correlation between the traditional
estimated VD and local number density of particles [panel (a)],
and the fourth order correlation between the VD and local number
density of particles [panel (b)] for model SCDM. The random
samples are the same as Fig. 3. The error bars are 1-$\sigma$
variance from 10 realizations.} \label{Fig13}
\end{figure}

\begin{figure}
\figurenum{14} \epsscale{0.6}
\plotfiddle{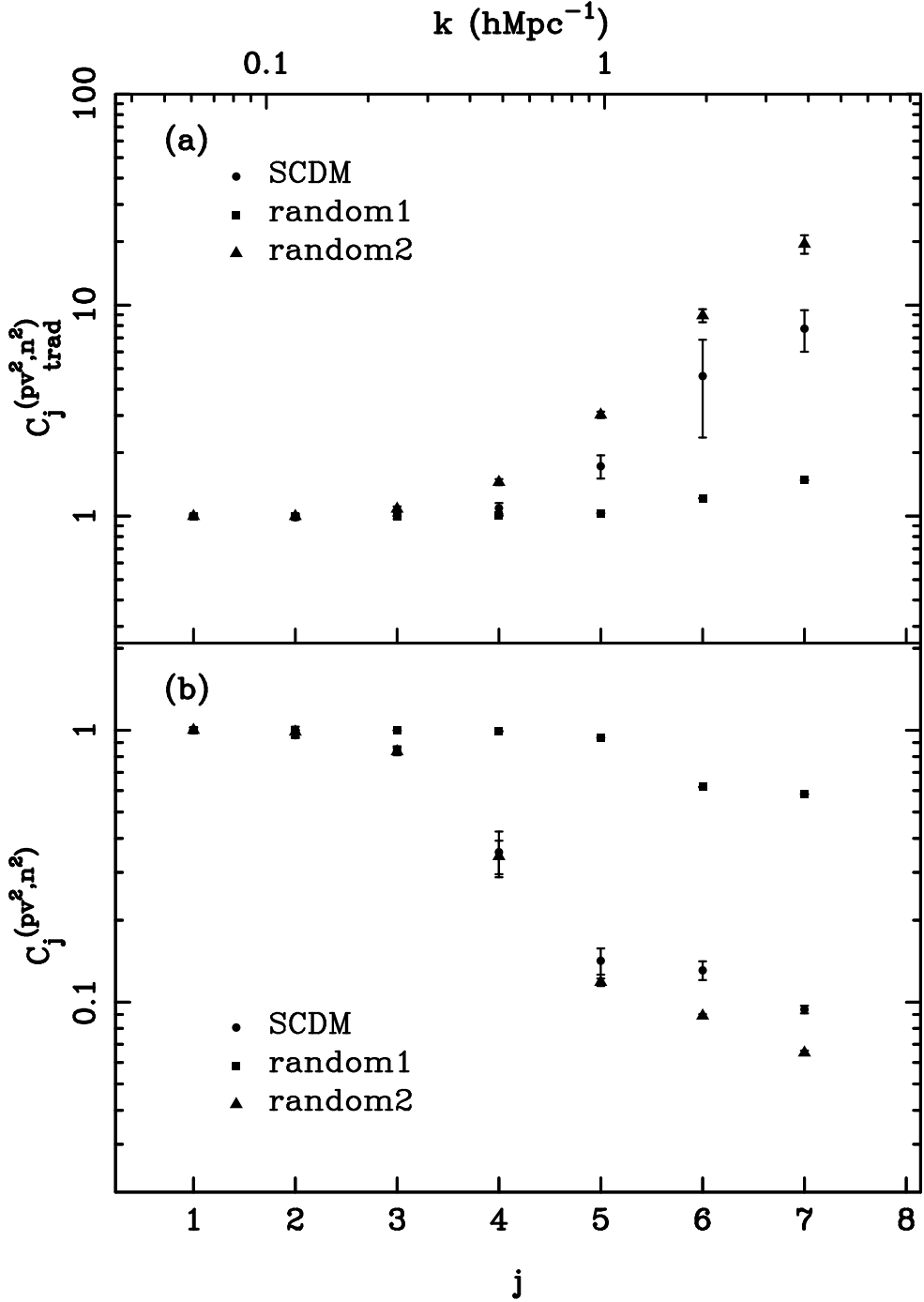}{20.cm}{0}{80}{80}{-230.}{-50.}
\caption[]{The fourth order correlations  $C^{(pv^2,n^2)}_{j \
trad}$ [eq.(57)] (upper panel), and $C^{(pv^2,n^2)}_j$ [eq.(58)]
(lower panel) for model SCDM.  The random samples are the same as
Fig. 3. The error bars are  1-$\sigma$ variance from 10
realizations.} \label{Fig14}
\end{figure}

\begin{figure}
\figurenum{15} \epsscale{0.6}
\plotfiddle{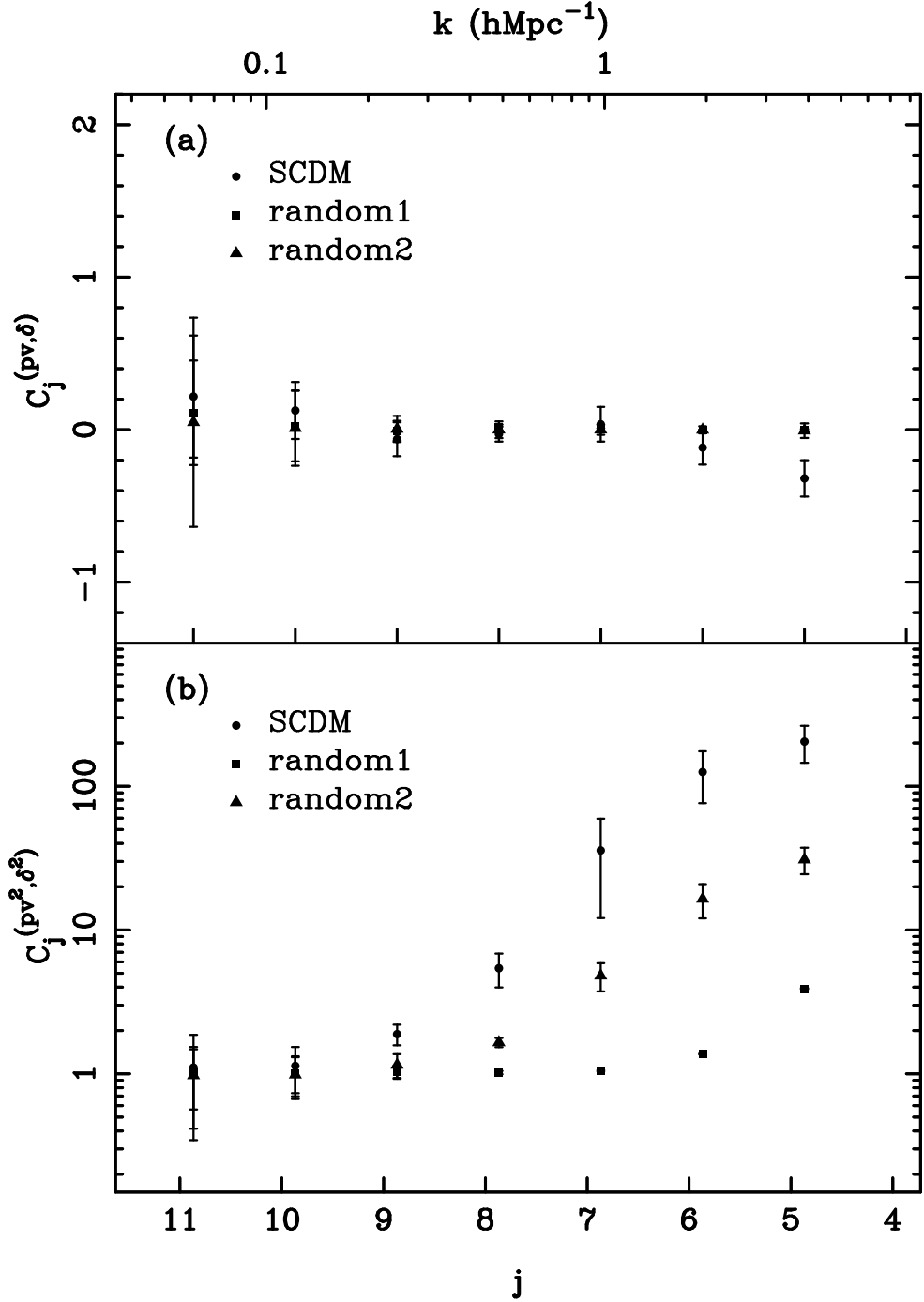}{20.cm}{0}{80}{80}{-230.}{-50.}
\caption[]{The second (upper panel) and fourth (lower panel) order
correlations between the PVD and local density fluctuation of
particles for model SCDM. The random samples are the same as Fig.
3. The error bars are 1-$\sigma$ variance from 10 realizations.}
\label{Fig215}
\end{figure}

\end{document}